\documentclass[twocolumn,aps,superscriptaddress,pre]{revtex4}

\usepackage{amsmath,amssymb,graphicx}
\usepackage{subfigure}
\usepackage{algorithmic}
\usepackage{enumerate}
\usepackage{times}
\usepackage{color}
\usepackage{rotating}
\usepackage{soul}
\definecolor{yblue}{rgb}{0.06, 0.3, 0.57}
\usepackage[pdftex]{hyperref}
\hypersetup{colorlinks=true,linkcolor=blue,citecolor=blue,urlcolor=blue}

\newcommand{\state}[1]{$\mathcal{S}_{#1}$}

\begin{document}

\title{Systematic vector solitary waves from their linear limits in one-dimensional $n$-component Bose-Einstein condensates}

\author{Wenlong Wang}
\email{wenlongcmp@scu.edu.cn}
\affiliation{College of Physics, Sichuan University, Chengdu 610065, China}

\begin{abstract}
We systematically construct a series of vector solitary waves in harmonically trapped one-dimensional three-, four-, and five-component Bose-Einstein condensates. These stationary states are continued in chemical potentials from the analytically tractable low-density linear limit of respective states, as independent linear quantum harmonic oscillator states, to the high-density nonlinear Thomas-Fermi regime. A systematic interpolation procedure is proposed to achieve this sequential continuation via a trajectory in the multi-dimensional space of the chemical potentials. The Bogolyubov-de Gennes (BdG) spectra analysis shows that all of the states considered herein can be fully stabilized in suitable chemical potential intervals in the Thomas-Fermi regime. Finally, we present some typical $SU(n)$-rotation-induced and driving-induced dynamics. This method can be extended to higher dimensions and shows significant promise for finding a wide range of solitary waves ahead.
\end{abstract}

\maketitle

\section{Introduction}
Solitons are ubiquitous nonlinear excitations in a wide range of dispersive and nonlinear waves, e.g., in Bose-Einstein condensates (BECs) \cite{becbook1,becbook2}, and nonlinear optics \cite{DSoptics}. 
Particularly, BECs have enjoyed considerable attention over the past decades, providing an ideal playground for investigating (single and multiple) solitonic structures, including studies of their generation, stability, interaction, instability, and associated dynamics and pattern formation \cite{Panos:book}. In the single-component one-dimensional setting, bright \cite{tomio} and dark solitons \cite{Dimitri:DS} are arguably the most fundamental structures in attractive and repulsive condensates, respectively. In higher dimensions, novel topological
structures bearing vorticity emerge such as vortices \cite{Alexander2001}, vortex rings \cite{fetter2,komineas}, and even knots \cite{PhysRevE.85.036306}. Moreover, extended dark solitonic structures, e.g., ring dark solitons in both two and three dimensions \cite{Panos:DSR,Wang:RDS} have also been considered.

In parallel, vector solitons in multicomponent settings are also fascinating. 
In one-dimensional two-component systems of repulsive interactions, the dark-bright structure has been a central point of theoretical and experimental efforts \cite{DBS1,Rajendran_2009,DDDB,DBcounterflow,Dong:MDB,DBtunneling,DBcollisions}. Here a bright component is trapped (and waveguided) by an effective potential of the dark soliton in the other component. It is important to note that a bright soliton cannot exist on its own in repulsive condensates, i.e., under self-defocusing nonlinearity. More exotic structures such as dark-dark \cite{Yan:DD,Wang:SO2}, and dark-antidark waves \cite{engels20} have also been found; see, e.g., recent works on magnetic solitons in both binary \cite{string,MS:20a} and spinor \cite{MS:20b} condensates.
Indeed, there has been a flurry of associated activities, as can be attested by
the recent works of~\cite{raman3,raman4}.
These solitary waves may naturally undergo dynamics not accessible in a single-component system \cite{Yan:DD,Lichen:DB,Lichen:RW}. 
While the two-component systems have been extensively studied \cite{revip}, there are far less studies on systems of three or even more components \cite{engels18,MS:20b}.
In this vein, it is especially relevant to highlight the fact that recent experimental
studies have rendered accessible a wide range of possibilities, including that
of suppressing the effect of spin-dependent interactions~\cite{Panos:BEC3C} and materializing
instantiations of the well-known Manakov model originally developed in nonlinear
optics~\cite{Manakov74}. While these recent developments have been predominantly
focusing on two- and three-component systems, it is natural to expect that generalizations
thereof to $F=2$ spin systems and up to five-component states are well within
reach~\cite{kawueda}.

There have been extensive theoretical efforts in generalizing the analytical techniques developed for the one-component system to multicomponent systems, e.g., the inverse scattering, the B\"{a}cklund transformation, the Darboux transformation, and the Hirota bilinear methods \cite{DT,Dressingmethod,Hirota,Lakshman}. The generalization is, however, frequently not straightforward and often extremely elaborate in its analytical form; see \cite{Lichen:DT} for a modified Darboux transformation and also a relevant discussion.
Moreover, these approaches are, by necessity, limited to the (integrable) one-dimensional homogeneous Manakov systems \cite{Manakov74} where the intra- and inter-species interactions are equal. This naturally poses the question of
developing methods that could be more straightforwardly generalized beyond the integrable realm to provide an understanding of the broad wealth
of nonlinear wave states that may become experimentally accessible in this
emerging setting of higher-component BEC systems.

In the non-integrable setting, e.g., in presence of a harmonic trap, finding stationary states and investigating their near-equilibrium dynamics and dynamical instabilities if any are especially interesting. The Bogolyubov-de Gennes (BdG) spectra are also natural to compute, encompassing much of the relevant dynamical information through the BdG eigenvalues and eigenvectors. Numerical methods can in principle find stationary solutions in a generic setting. Recently, a deflation method \cite{Panos:DC1,Panos:DC2,Panos:DC3} was studied which runs iteratively at fixed chemical potentials, and the solver is designed such that a new solution, if converges properly, avoids already found ones by properly modifying the stationary-state equation to solve. This method has found a remarkable series of solutions in both one- and two-component systems. However, the exploration of the solution space is not controlled, and the method inevitably becomes increasingly expensive as more states are added to the list of solutions.

An alternative semi-analytical method is to construct solitary waves from the known linear limit in a suitable, e.g., harmonic potential using the (chemical potential) parametric continuation \cite{PK:var,Wang:OD,Wang:DBS,Wang:DD}. In the linear limit, the nonlinear term is negligible and the linear problem is fully solvable as different components decouple into independent quantum harmonic oscillators. Perturbation analysis suggests that a low-density linear state can be continued in chemical potentials to a weakly nonlinear one, and thereafter to a highly nonlinear state in the high-density Thomas-Fermi regime, i.e., a series of solutions can be constructed interpolating the two limits. The spirit of the method is therefore to take advantage of the analytically tractable linear limit by first turning off the nonlinearity and then gradually adding it back. In fact, the recent three-dimensional deflation study also partially employed this idea \cite{Panos:DC3}, showing the significance of the method. This method has recently been successfully applied to the one-dimensional two-component system, focusing instead on solitonic beating patterns following a unitary rotation or mixing of the different components, along with two case examples in the three-component setting \cite{Wang:DD}. It is also worth mentioning that while these states are constructed in the harmonic potential, further continuation to other potentials, e.g., by interpolating between two different potentials is possible, showing the flexibility of the method.

The main purpose of the present work is to systemically construct solitary waves from the linear limit for a general $n$-component system, motivated in particular,
by the above discussion and recent
experimental implementation of the three-component Manakov model \cite{Panos:BEC3C}.
The availability of $F=1$ and $F=2$ three- and five-component systems \cite{kawueda}, respectively, prompts us to
illustrate the method using $n=3, 4$, and $5$. An additional motivation is that it is sensible to demonstrate the effectiveness of the method in 1+1 dimension before further extending it to higher dimensions, where there is a ``degenerate state problem''. Our approach is so far successful, and a large series of states of increasing complexity are constructed. Their stability properties are also considered. Despite the expectation that, in principle, it typically gets increasingly harder to stabilize more complex states, bearing a growing number of the so-called negative energy
modes \cite{Panos:book}, it is remarkable that all of the states considered herein can be properly stabilized in suitable chemical potential intervals as they approach the Thomas-Fermi limit. 
Finally, some typical dynamics are illustrated. The states should be able to access a  rich set of dynamical evolution scenarios, considering their complexity. Indeed, our direct numerical simulations confirm this expectation.
Here, we only present a few prototypical proof-of-principle examples for clarity. Specifically, we focus on two types of dynamics: $SU(n)$-rotation-induced, and driving-induced beating dynamics. Both periodic and aperiodic dynamics are accessible and are illustrated.

The presentation is organized as follows. In Sec.~\ref{setup}, we introduce 
the model, the numerical setup, and the method of constructing vector solitary waves from the linear limit.
Next, we present our results in Sec.~\ref{results}. 
Finally, our conclusions and a number of open
problems for future consideration are given in Sec.~\ref{conclusion}, while the
Appendix discusses the formulation of the BdG analysis in the general $n$-component
case.

\section{Model and methods}
\label{setup}

We first present the mean-field Gross-Pitaevskii equation and the $SU(n)$ symmetry for $n$-component condensates with Manakov interactions, and the numerical methods used for finding stationary states, stability analysis, and dynamics. Then we discuss the method of constructing stationary vector solitary waves from the linear limit using the chemical potential continuation, and the scaling of the number of solitons with the principle or maximum quantum number.

\subsection{Computational setup}

In the framework of the lowest-order mean-field theory, and for sufficiently low temperatures, the dynamics of one-dimensional $n$-component repulsive BECs, 
confined in a time-independent trap $V$, is described by the following coupled
dimensionless Gross-Pitaevskii equation (GPE) \cite{Panos:book,Wang:DD}:
\begin{eqnarray}
i \frac{\partial \psi_j}{\partial t} = -\frac{1}{2} \psi_{jxx}+V \psi_j + \left(\sum_{k=1}^n g_{jk}| \psi_k |^2\right) \psi_j,
\label{GPE}
\end{eqnarray}
where $\psi_j, j=1, 2, ..., n$ are $n$ complex scalar macroscopic wavefunctions. We focus here on the Manakov system of repulsive interactions $g_{ij}=1$ for simplicity, but the method is not limited to this constraint. While earlier spinor condensates \cite{engels18,MS:20b} also contain spin-dependent interactions \cite{kawueda,stampueda}, recently spinor condensates with Manakov interactions become available \cite{Panos:BEC3C}. Moreover, the work is also partially motivated by multicomponent nonlinear optical problems \cite{Park:Opticalsolitons}. Nevertheless, we expect that the solitary waves considered herein should be relevant more broadly, as spin-dependent interactions are typically small.

The condensates, unless otherwise specified, are confined in a harmonic trap of the form: 
\begin{equation}
V=\frac{1}{2} \omega^2 x^2,
\label{potential}
\end{equation}
where the trapping frequency is set to $\omega=1$ by scaling without loss of generality. Stationary states of the form:
\begin{eqnarray}
\psi_j(x,t) = \psi^0_j(x)e^{-i\mu_jt}
\label{ss}
\end{eqnarray}
lead to $n$ coupled stationary equations:
\begin{eqnarray}
\label{SS1}
-\frac{1}{2} \psi^0_{jxx}+V \psi^0_j +\left(\sum_{k=1}^n g_{jk}| \psi_k^0 |^2\right) \psi^0_j = \mu_j \psi^0_j,
\end{eqnarray}
where $\mu_j$ is the chemical potential of the $j$th component.

Equation~(\ref{GPE}) has $n$ $U(1)$ symmetries, i.e., if $(\psi_1, ..., \psi_n)^T$ is a solution, then $(\psi_1 e^{i\theta_1}, ..., \psi_ne^{i\theta_n})^T$ is also a solution, where $\{\theta_j\}$ are real numbers. In the Manakov case, there is an additional $SU(n)$ symmetry. It is straightforward to show that $(\psi_1', ..., \psi_n')^T=U(\psi_1, ..., \psi_n)^T$
is also a solution if $U$ is unitary, $UU^{\dagger}=\mathbb{I}$. Note that the total density profile is invariant upon the rotation, i.e., $\sum_j|\psi_j'|^2 = \sum_j|\psi_j|^2$. Because a stationary state typically has different chemical potentials for each component, the mixed states after rotation are typically dynamical states.


Next, we present the numerical details. A stationary state, given a proper initial guess as detailed below, is computed using a finite element method for the discretization of space and the iterative Newton's method towards convergence. The linear oscillator states are used as the initial guess for a stationary state near but not at the linear limit, and the converged state is then served as the initial guess for the next nearby chemical potentials and so on. We use a linear ``trajectory'' for simplicity in the multidimensional $\vec{\mu}=(\mu_1, ..., \mu_n)^T$ parameter space, i.e., given the linear limit chemical potentials $\vec{\mu}_i$ and the final (chosen) chemical potentials $\vec{\mu}_f$, the trajectory is given by $\vec{\mu}=\vec{\mu}_i+\epsilon(\vec{\mu}_f-\vec{\mu}_i)$, where $\epsilon \in (0,1]$ is a parameter interpolating the two points. If $\vec{\mu}_i$ and $\vec{\mu}_f$ are given, it is sufficient to specify a point by either $\epsilon$ or any of the chemical potentials, e.g., $\mu_1$ as we shall do below. To find the first weakly-coupled stationary state, we set the initial chemical potentials approximately $O(0.01)$ away from the linear ones. The finial point is empirically chosen such that the maximum densities do not vary significantly between adjacent components, while keeping the order of the chemical potentials unchanged, i.e., we keep $\mu_1>...>\mu_n$. In fact, the continuation of states is pretty robust and straightforward in the one-dimensional setting, the careful selection of the final chemical potentials is to help finding spectrally stable states.

The BdG stability spectrum is computed for each stationary state found along the trajectory. The BdG analysis is a linear stability analysis of a stationary state, and it is described in the Appendix for clarity. For each stationary state, we compute the first $100$ low-lying eigenvalues in magnitude and the eigenvectors. The eigenvalues are generally complex $\lambda = \lambda_r + i\lambda_i$. If there are modes with $\lambda_r>0$, the state is dynamical unstable with respect to perturbations. On the other hand if all the eigenvalues are entirely imaginary, the state is robust and dynamically stable. As the states get more complex, the number of unstable modes stemming from the linear limit tends to increase, which requires higher chemical potentials to suppress the instabilities. This consequently requires both a larger domain and a finer spacing for a more complex state, i.e., studying a more complex state is more computationally expensive.

We select stable states, i.e., at suitable chemical potentials where the BdG eigenvalues are all imaginary for $SU(n)$-induced \cite{Yan:DD} and driving-induced \cite{Lichen:DB} dynamics. For the former, different components are mixed producing either periodic or aperiodic beating patterns depending on the specific chemical potentials. For the latter, we apply a constant driving force $F>0$ along the negative $x$-axis to one component, i.e., the component experiences an additional linear potential of $V_D=Fx$. Typically, we drive the ``bright'' component, which has no node, producing approximately periodic orbitals \cite{Lichen:DB}. Our dynamics are integrated using the regular fourth-order Runge-Kutta method.

\subsection{Construct vector solitary waves from the linear limit}
\label{ll}

The idea of constructing solitary waves from the linear limit is extremely simple but effective. For completeness, we start from the one-component setting. In this case, each harmonic oscillator state $|n_1\rangle$ with the chemical potential or eigenvalue $n_1+1/2$ can be continued to the Thomas-Fermi regime containing $n_1$ dark solitons \cite{PK:var}. In this process, the number of particles $N$ is approximately $0$ near the linear limit, and then it grows as the chemical potential is increased. For example, the ground state has a linear limit at $\mu_1=0.5$ as a faint Gaussian function. As the chemical potential increases, it becomes the Thomas-Fermi ground state. The first excited state has a linear limit at $\mu_1=1.5$, and in a similar process it turns into a single dark soliton state embedded in the Thomas-Fermi sea. The nonlinear wave stemming from the linear state $|n_1\rangle$ contains a total of $n_1$ dark solitons, which can be conveniently labelled as \state{n_1}. Here, $\mathcal{S}$ stands for state or soliton.

\begin{table}
\caption{
Number of states with distinct quantum numbers from the linear limit. There are a total of $C^{n-1}_{n_1}=\frac{n_1!}{(n-1)!(n_1-n+1)!}$ states in the family $n_1$ (the maximum quantum number) for the $n$-component system. Note that the asymptotic growth speed is increasingly rapid as the number of component $n$ grows.
\label{table}
}
\begin{tabular*}{\columnwidth}{@{\extracolsep{\fill}} l c c c c c c c c c c r}
\hline
\hline
$n$  &$n_1=0$ &$1$  &$2$ &$3$ &$4$ &$5$ &$6$ &$7$ &$8$ &$9$ &$10$ \\
\hline
$1$  &$1$ &$1$ &$1$ &$1$ &$1$ &$1$ &$1$ &$1$ &$1$ &$1$ &$1$ \\
$2$  &$-$ &$1$ &$2$ &$3$ &$4$ &$5$ &$6$ &$7$ &$8$ &$9$ &$10$ \\
$3$  &$-$ &$-$ &$1$ &$3$ &$6$ &$10$ &$15$ &$21$ &$28$ &$36$ &$45$ \\
$4$  &$-$ &$-$ &$-$ &$1$ &$4$ &$10$ &$20$ &$35$ &$56$ &$84$ &$120$ \\
$5$  &$-$ &$-$ &$-$ &$-$ &$1$ &$5$ &$15$ &$35$ &$70$ &$126$ &$210$ \\
\hline
\hline
\end{tabular*}
\end{table} 

For a two-component system, the linear limit has two quantum numbers from the two independent harmonic oscillators $|n_1, n_2\rangle$ \cite{Wang:OD,Wang:DD}. The state has its linear limit at $(\mu_1, \mu_2)=(n_1+1/2, n_2+1/2)$. We focus here on states $n_1>n_2 \geq 0$, as it is not hard to prove that $\langle \psi_i^0 | \psi_j^0 \rangle (\mu_i-\mu_j)=0$, i.e., if two states are not orthogonal, they must have the same chemical potentials. For example, the state $|1,1\rangle$ can indeed be continued to the stationary dark-dark soliton, but because the two components must have the same chemical potential, the two profiles are in fact identical up to a scaling factor. This state is therefore somewhat trivial in the sense that it can be obtained by splitting the corresponding single dark soliton state of the one-component system. Note that if $\psi^0$ is a one-component stationary state, then $(c_1\psi^0, c_2\psi^0)^T$ is a stationary state of the two-component Manakov system if $|c_1|^2+|c_2|^2=1$. Such splitting can be readily generalized, if we have an $n$-component stationary state, we can split any of the component in the same way to get an $(n+1)$-component stationary state. Therefore, we focus here on irreducible states where all the pertinent quantum numbers are distinct. The two-component system has recently been systematically explored in \cite{Wang:DD}. The low-lying states are \state{10}, \state{20}, and \state{21} corresponding to the well-known single dark-bright, the in-phase two dark-bright, and the out-of-phase two dark-bright structures, respectively.

The procedure can be generalized to $n$-component systems. Specifically, we can continue the harmonic oscillator state $|n_1,...,n_n\rangle$ to the solitary wave $\mathcal{S}_{n_1,...,n_n}$, where again $n_1>...>n_n \geq 0$. In this work, we explore the three-component setting systematically, and further study some prototypical low-lying states in four- and five-component systems. It should be noted that the number of states grows very rapidly with the increasing principle quantum number $n_1$ in multicomponent systems. It is straightforward to show that the number of states in the family $n_1$ for the $n$-component system is given by $C^{n-1}_{n_1}=\frac{n_1!}{(n-1)!(n_1-n+1)!} \sim \frac{n_1^{n-1}}{(n-1)!}$ as $n_1 \rightarrow \infty$. The asymptotic growth with $n_1$ is therefore increasingly rapid as $n$ increases, it is constant for $n=1$, linear for $n=2$, quadratic for $n=3$, and so on. The number of low-lying states are summarized in Table~\ref{table}. In this work, we exhaust all the states in the three-component system up to $n_1=4$, and study some typical higher-lying states up to $n_1=10$. In four- and five-component systems, we study the respective $6$ lowest-lying states. We shall see below that these state profiles are already quite complex.

\section{Numerical results}
\label{results}

\subsection{Vector solitary waves from the linear limit}
\begin{figure*}
\subfigure[]{\includegraphics[width=0.24\textwidth]{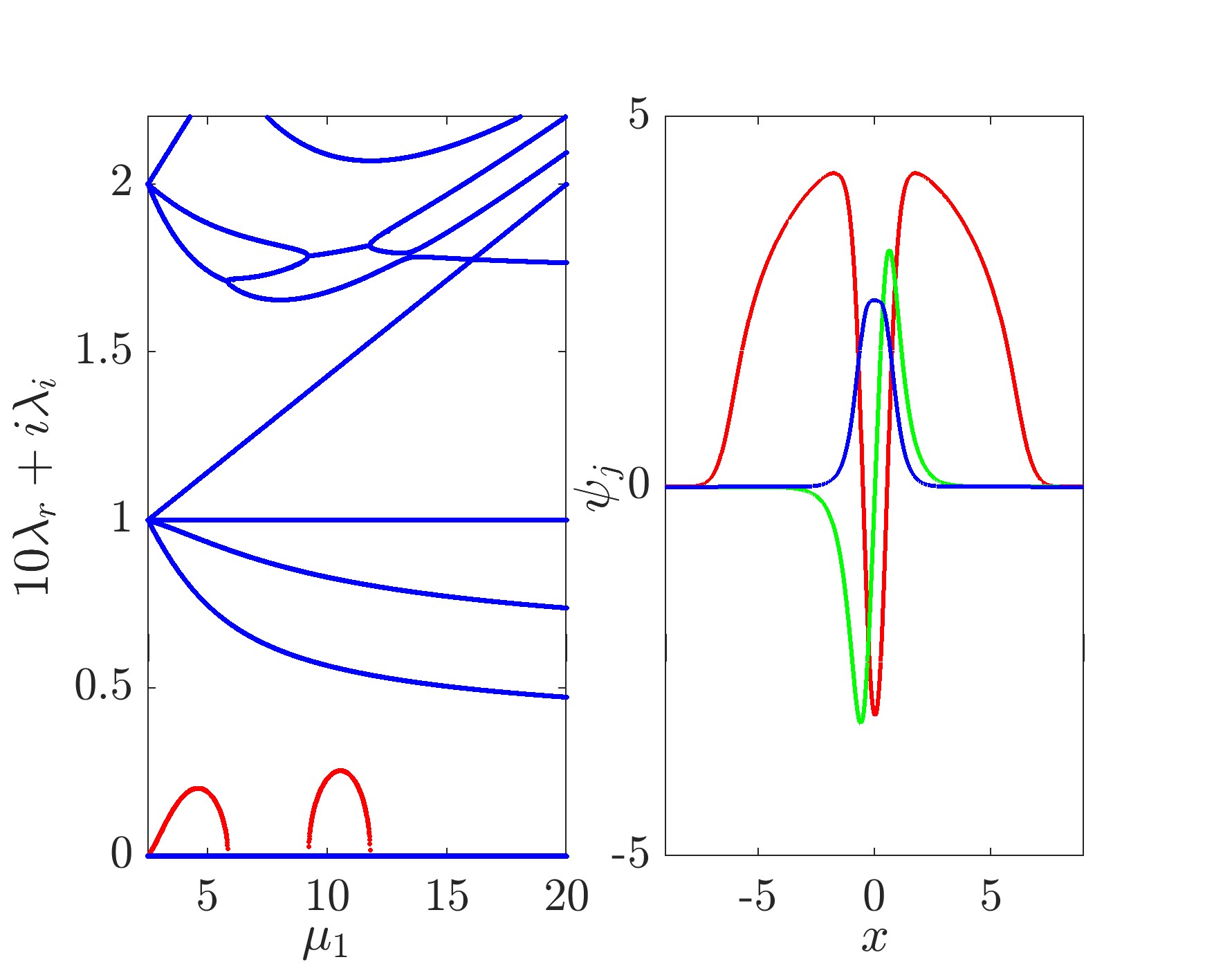}}
\subfigure[]{\includegraphics[width=0.24\textwidth]{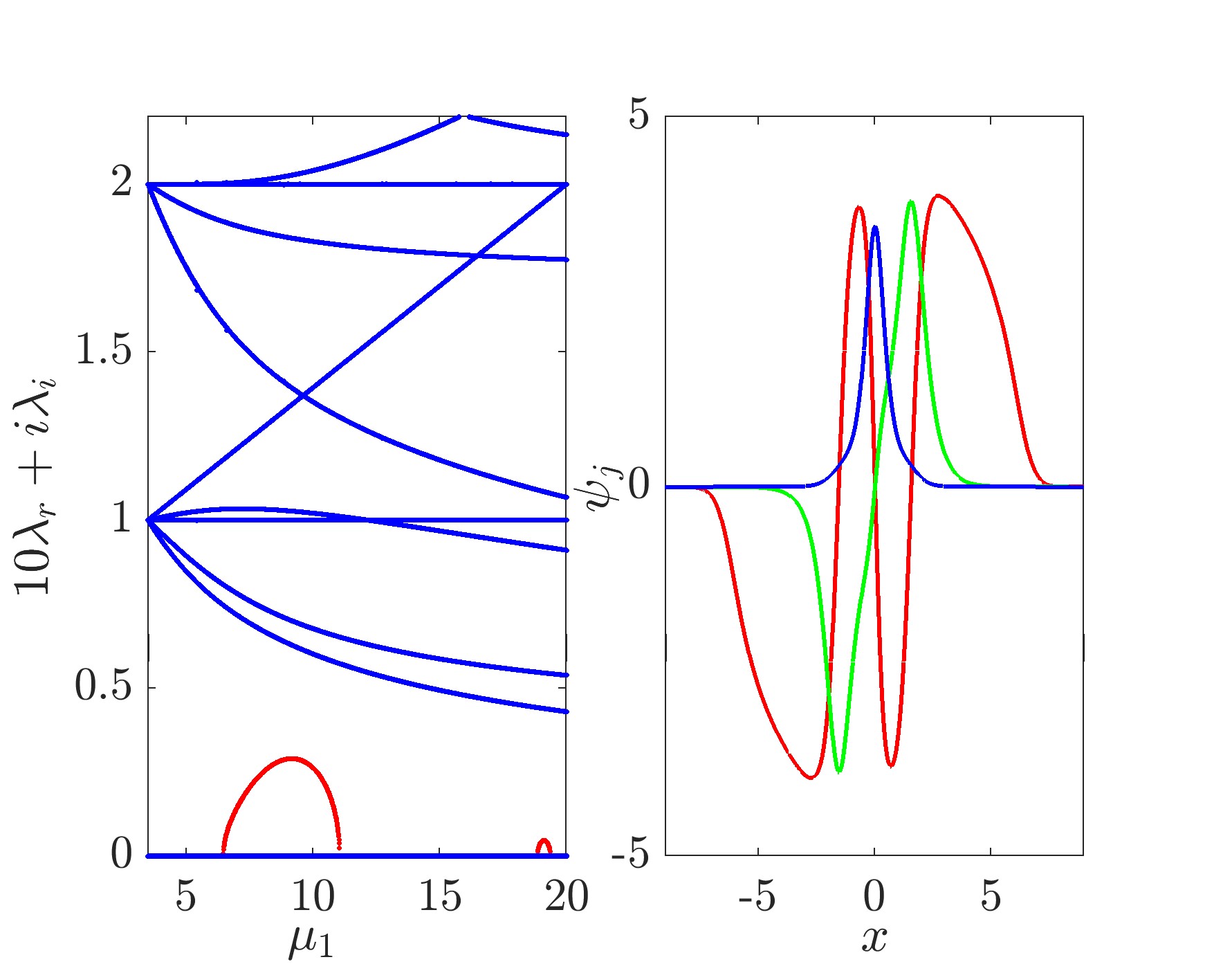}}
\subfigure[]{\includegraphics[width=0.24\textwidth]{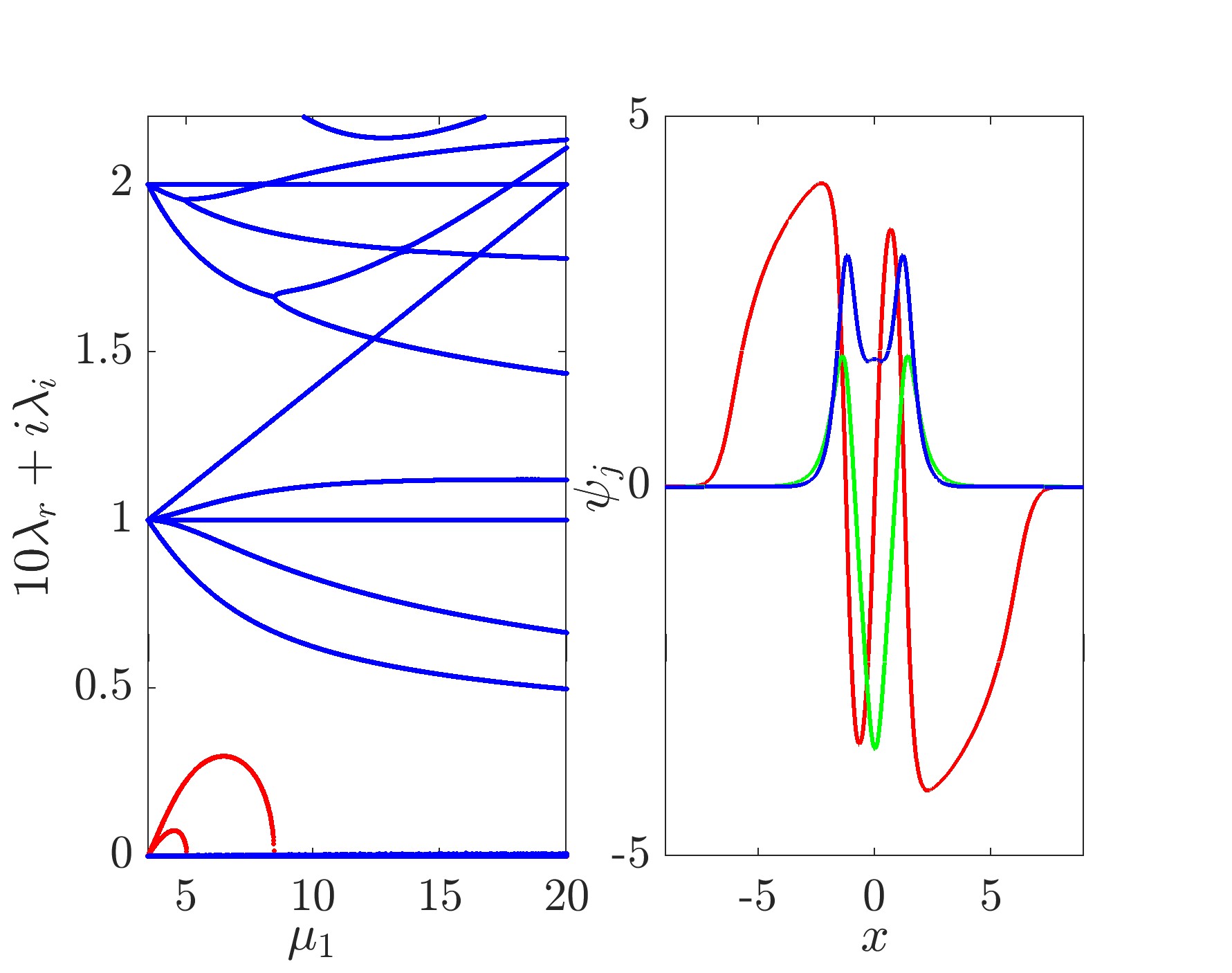}}
\subfigure[]{\includegraphics[width=0.24\textwidth]{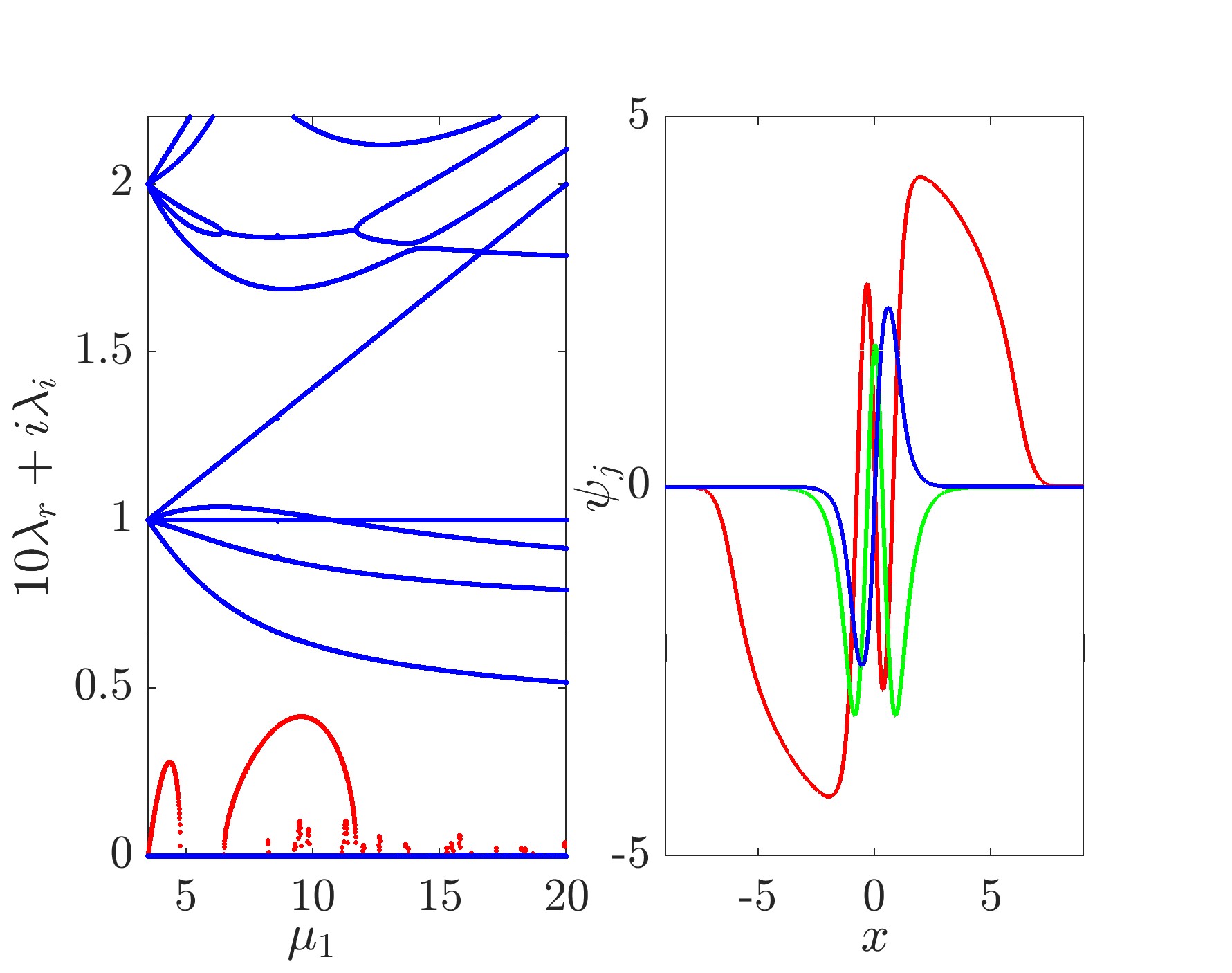}}
\subfigure[]{\includegraphics[width=0.24\textwidth]{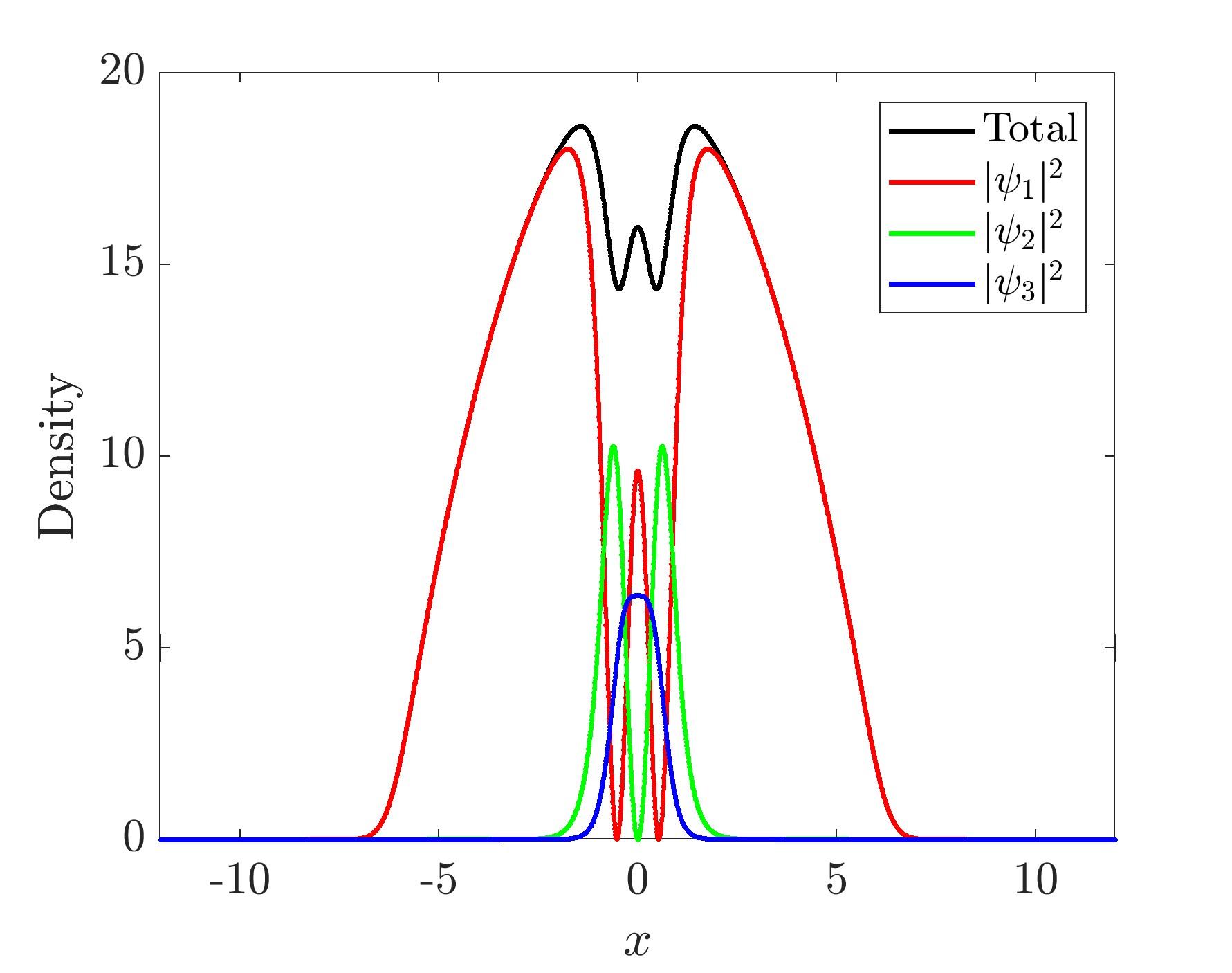}}
\subfigure[]{\includegraphics[width=0.24\textwidth]{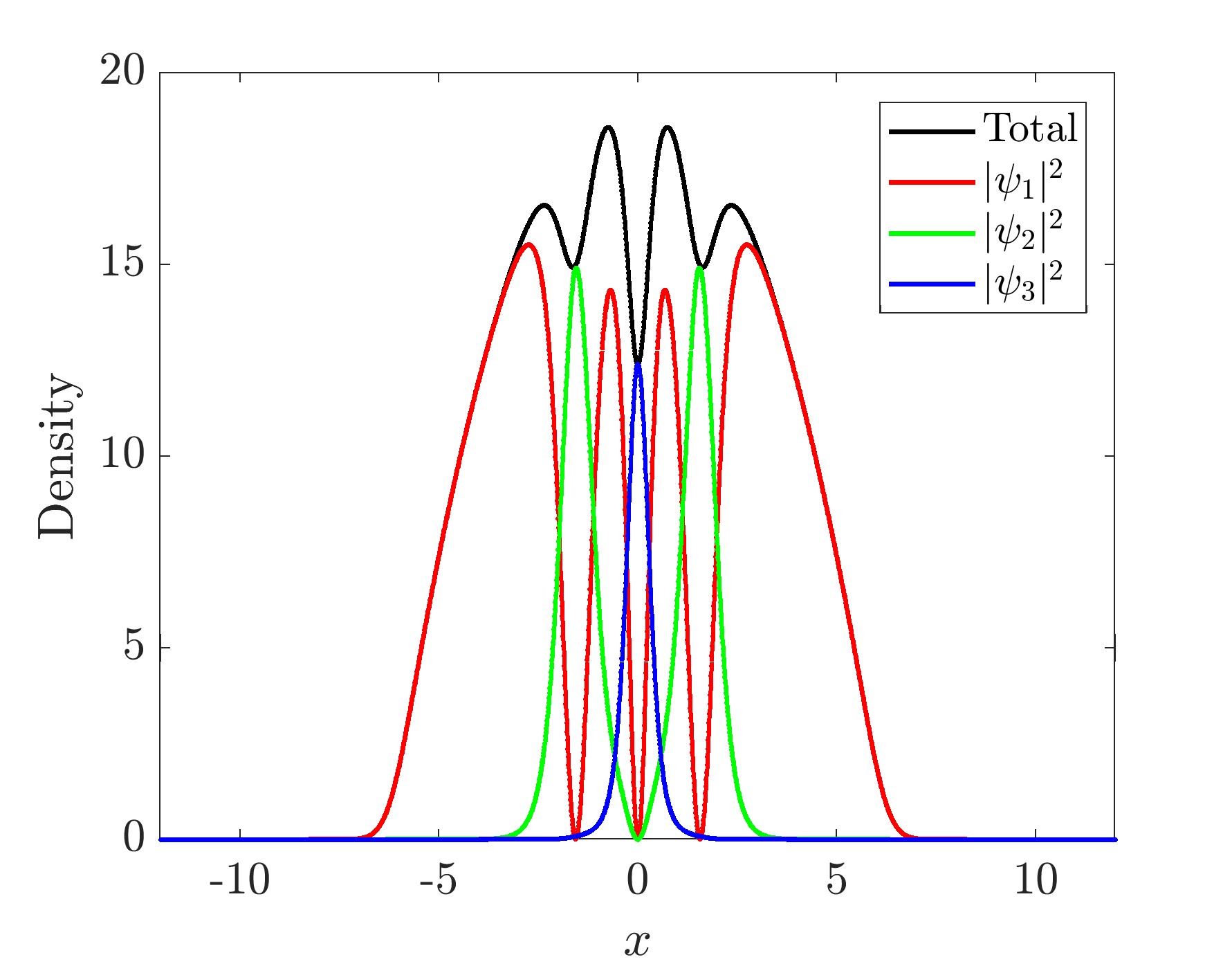}}
\subfigure[]{\includegraphics[width=0.24\textwidth]{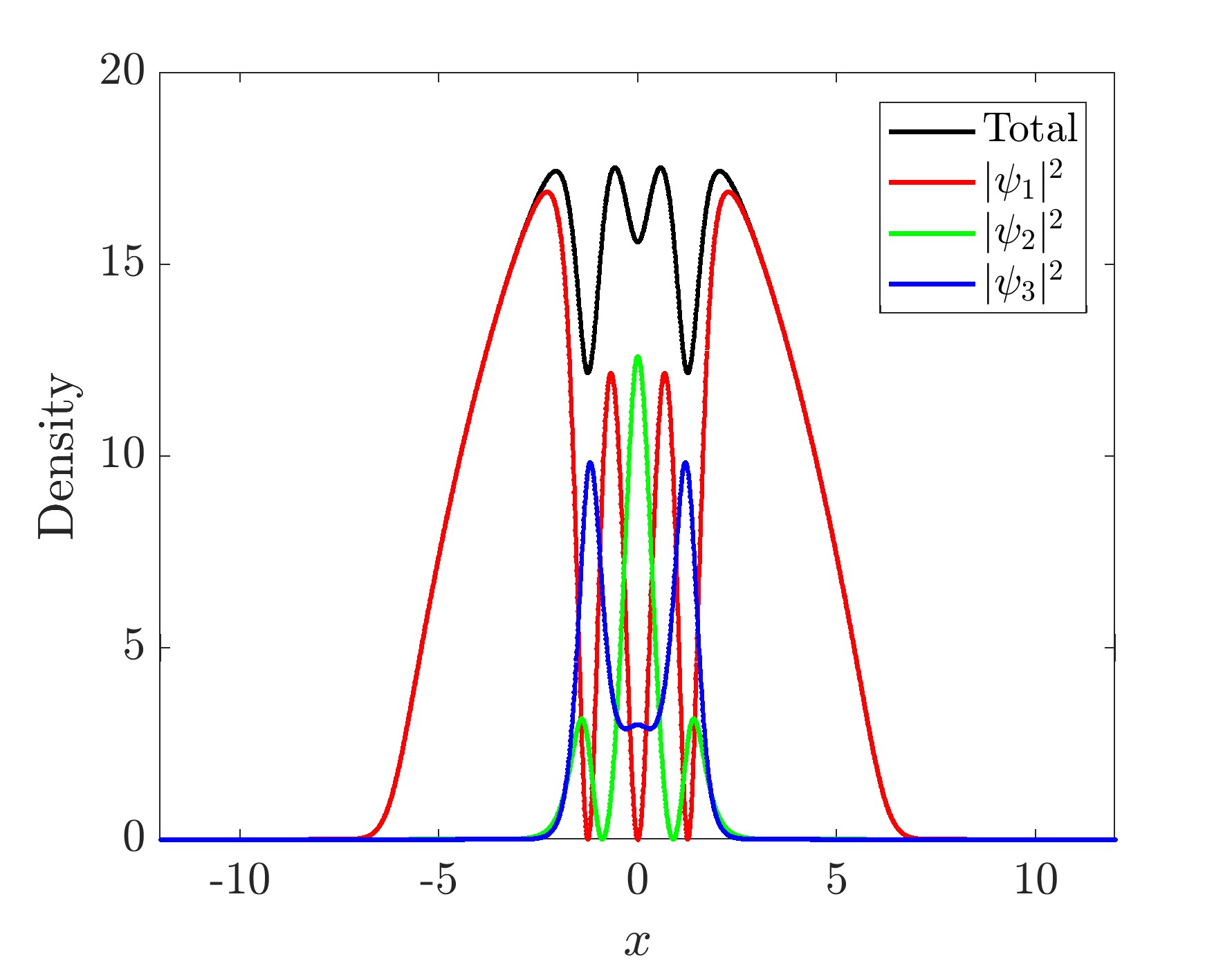}}
\subfigure[]{\includegraphics[width=0.24\textwidth]{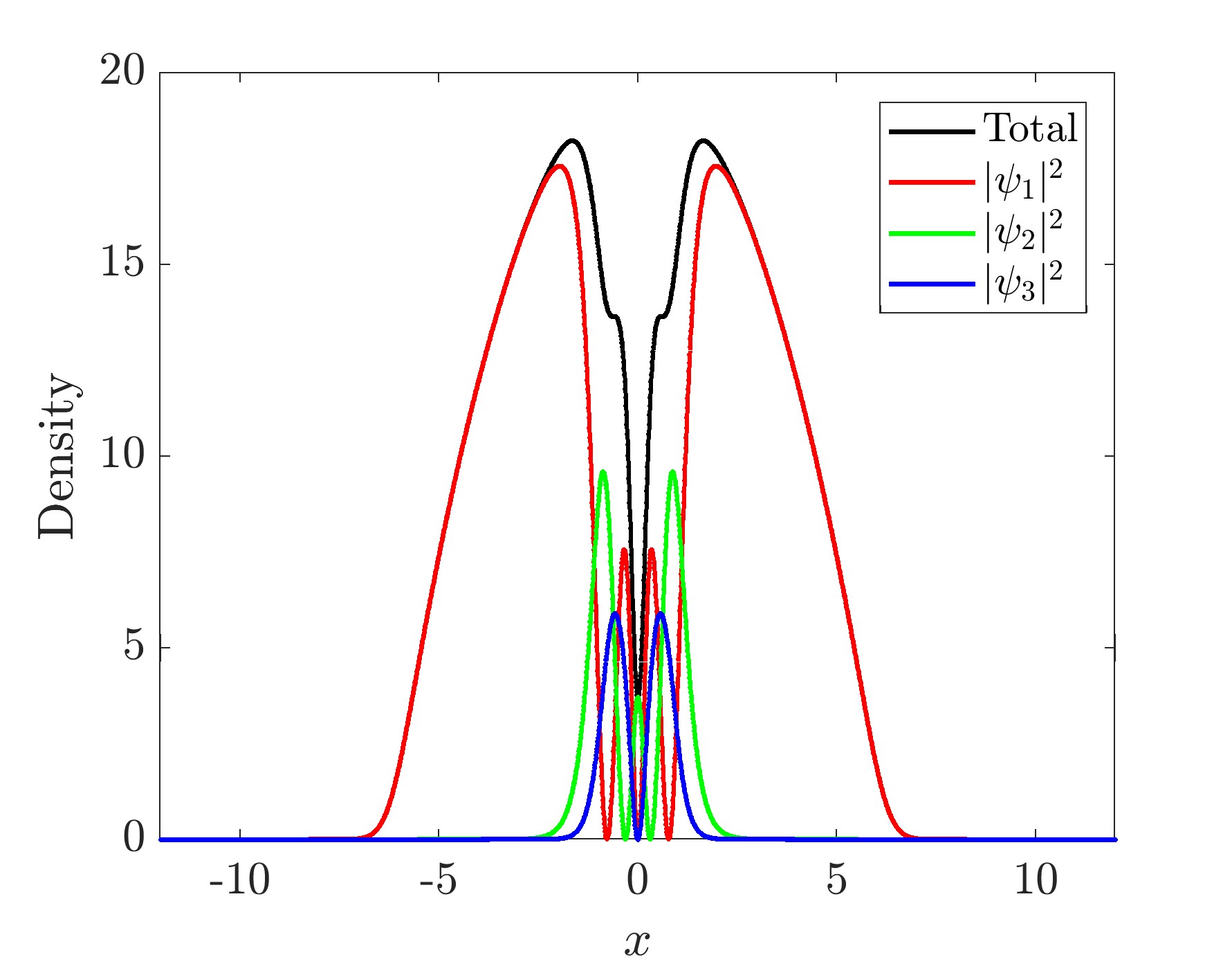}}
\caption{
\textit{Top panels}: Left of (a): BdG spectrum $\lambda$ of the \state{210} state along a linear trajectory from the linear limit $(2.5,1.5,0.5$) to a typical large-density limit $(20,18,16)$ in the $(\mu_1,\mu_2,\mu_3)$ parameter space, only $\mu_1$ is shown but all chemical potentials are varied due to the constraint of the prescribed linear trajectory. Red and blue points are for the (unstable) real and (stable) imaginary parts of the eigenvalues, respectively. Note that the real part of the eigenvalues is enlarged by a factor of $10$ for ease of visualization, i.e., the maximum growth rate is approximately $0.3/10=0.03$. Right of (a): A typical configuration at $\mu_1=20$ is depicted, the three components are in turn plotted in red, green, and blue. The other panels (b-d) are the same but for the states \state{310}, \state{320}, and \state{321}, respectively. \textit{Bottom panels}: The total density profile and the density profiles of each component of the states depicted above.
}
\label{S210}
\end{figure*}

We start from the three-component system, the first few low-lying states and their BdG spectra are depicted in Fig.~\ref{S210}. The first observation is that these states exist, and all of them contain certain unstable modes (the red curves of the spectra are for the real part of the eigenvalues $\lambda_r$ and instabilities, the blue curves are for the imaginary part of the eigenvalues $\lambda_i$ and stable modes). By contrast, the one-component dark soliton and the two-component dark-bright soliton and even the in-phase two dark-bright solitons appear to be very robust structures \cite{Wang:DD}. Similarly, the stability tends to be improved as a state moves towards the Thomas-Fermi limit, and there are suitable chemical potential intervals where these solitary waves are fully stable. It should be noted that the instabilities are very weak though for these low-lying structures, the real part of the eigenvalues is enlarged by a factor of $10$ in Fig.~\ref{S210} for ease of visualization, i.e., the maximum growth rate is only about $0.3/10=0.03$; cf. the trapping frequency $\omega=1$.

In order to gain more insight on the structures, we examine the total density profile (as an effective density potential) and the density profiles of the trapping first component and the trapped second and third components, shown in the bottom panels of Fig.~\ref{S210}. Interestingly, the total density profiles are quite different among the structures, they also do not exhibit a Thomas-Fermi structure, but local density minimums are found in all cases. In state \state{210}, there is a double well potential structure, the two peaks of the second component concentrate at the two wells while the third component is trapped at the center by the edges of the double well. Note that each component is also trapped by the external harmonic potential. The state \state{310} has three density wells, the two peaks of the second component occupy the side wells while the peak of the third component concentrates in the central well. The state \state{320} also has three wells, but the side ones are deeper. Here, the second component has three peaks sit in the three wells, the central peak is the most prominent. The third component has two prominent peaks concentrated in the two side wells. It also has a finite weight with a barely peak structure in the central well. The state \state{321} has a single funnel-like potential trapping altogether both the second and third components with three and two peaks, respectively. These five peaks of the two components are organized alternatively, with the side peaks larger and the central peaks smaller.

\begin{figure*}
\subfigure[]{\includegraphics[width=0.33\textwidth]{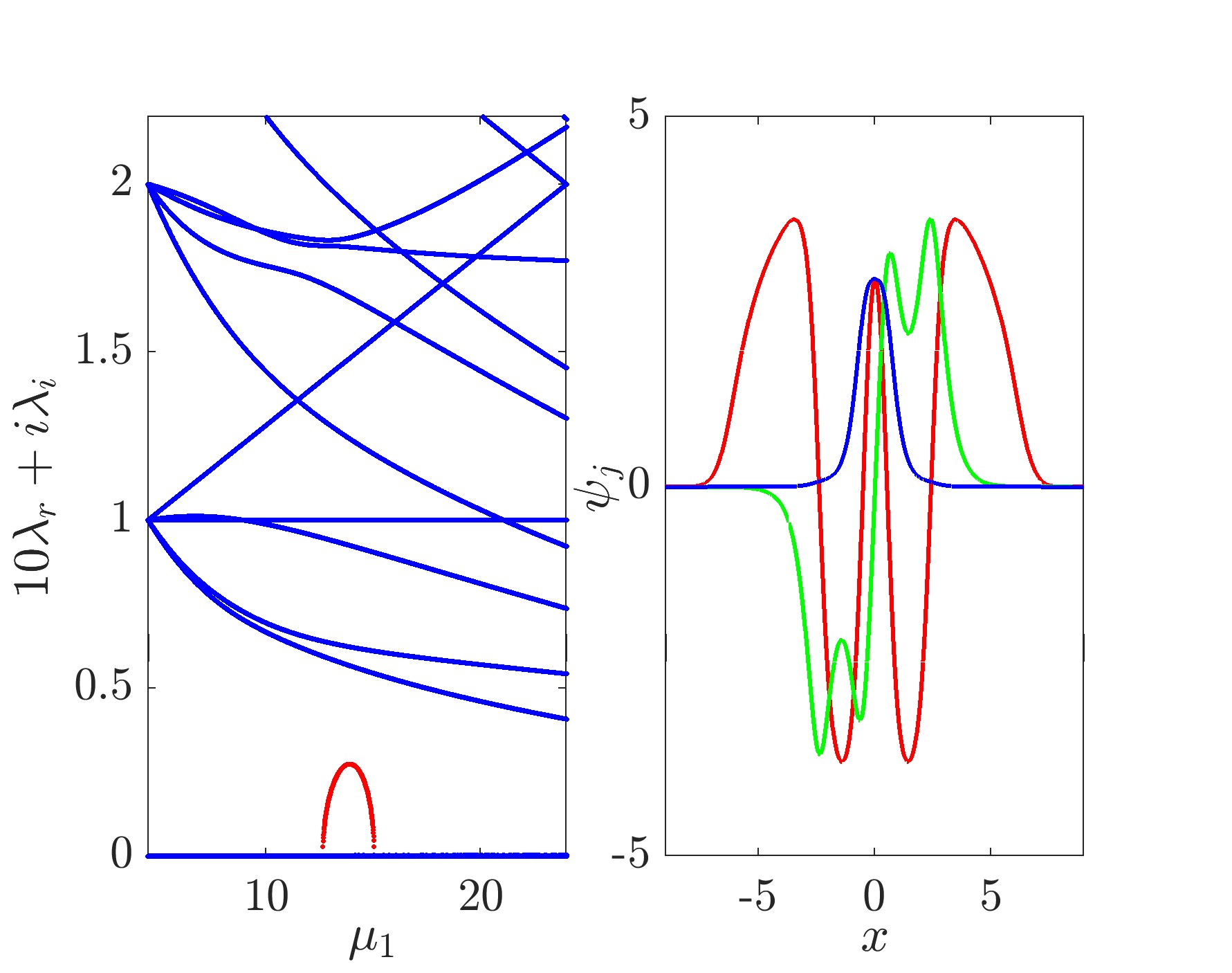}}
\subfigure[]{\includegraphics[width=0.33\textwidth]{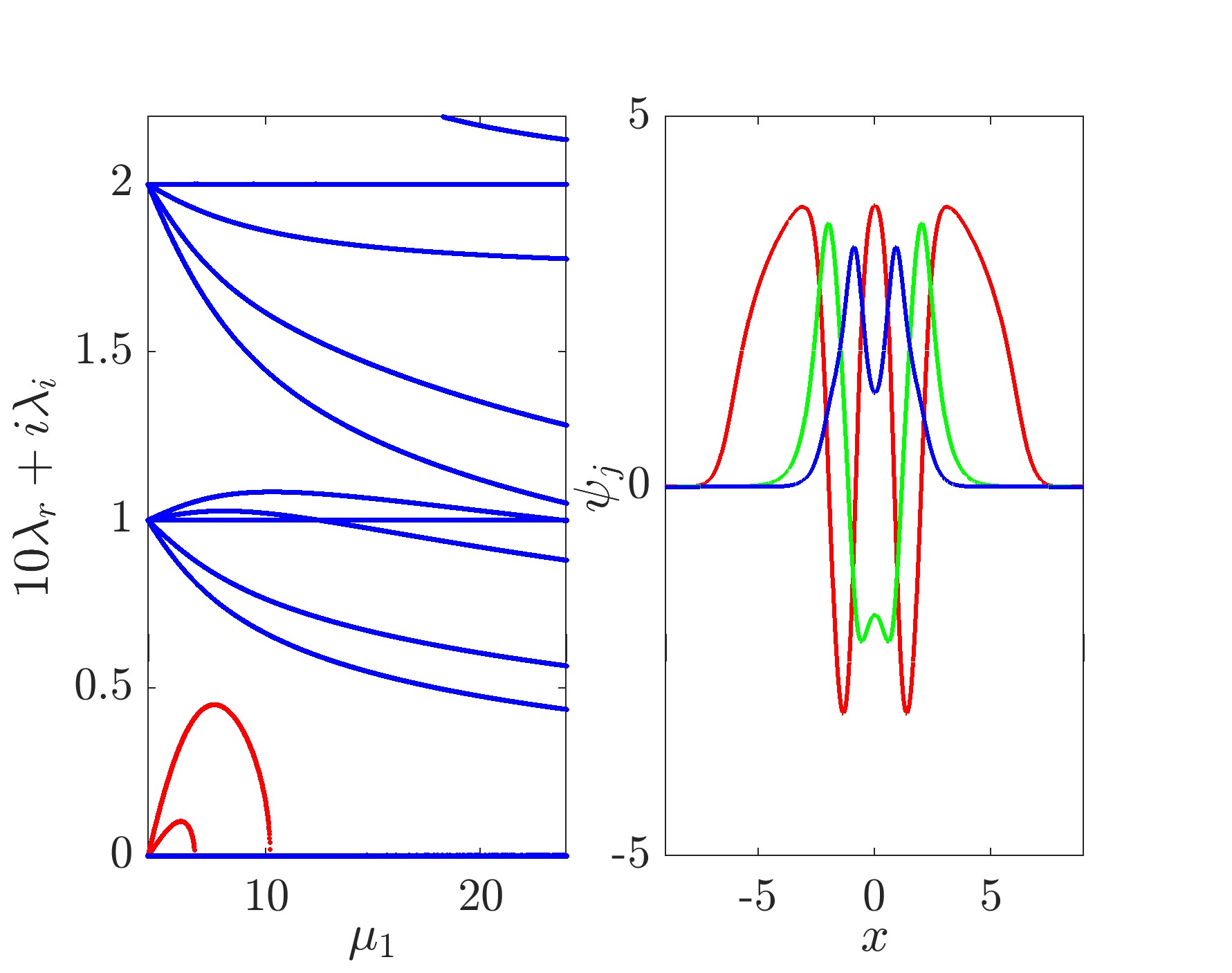}}
\subfigure[]{\includegraphics[width=0.33\textwidth]{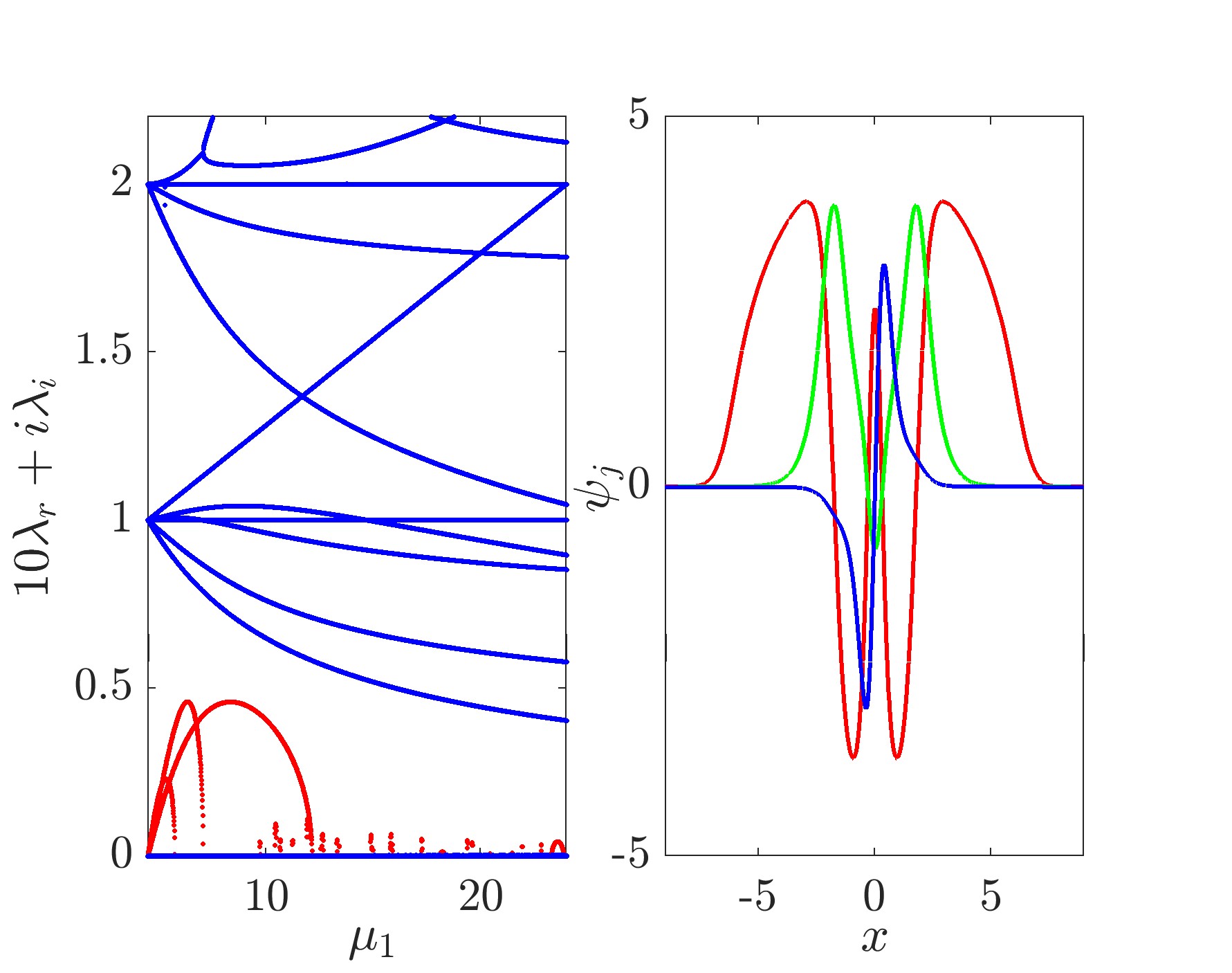}}
\subfigure[]{\includegraphics[width=0.33\textwidth]{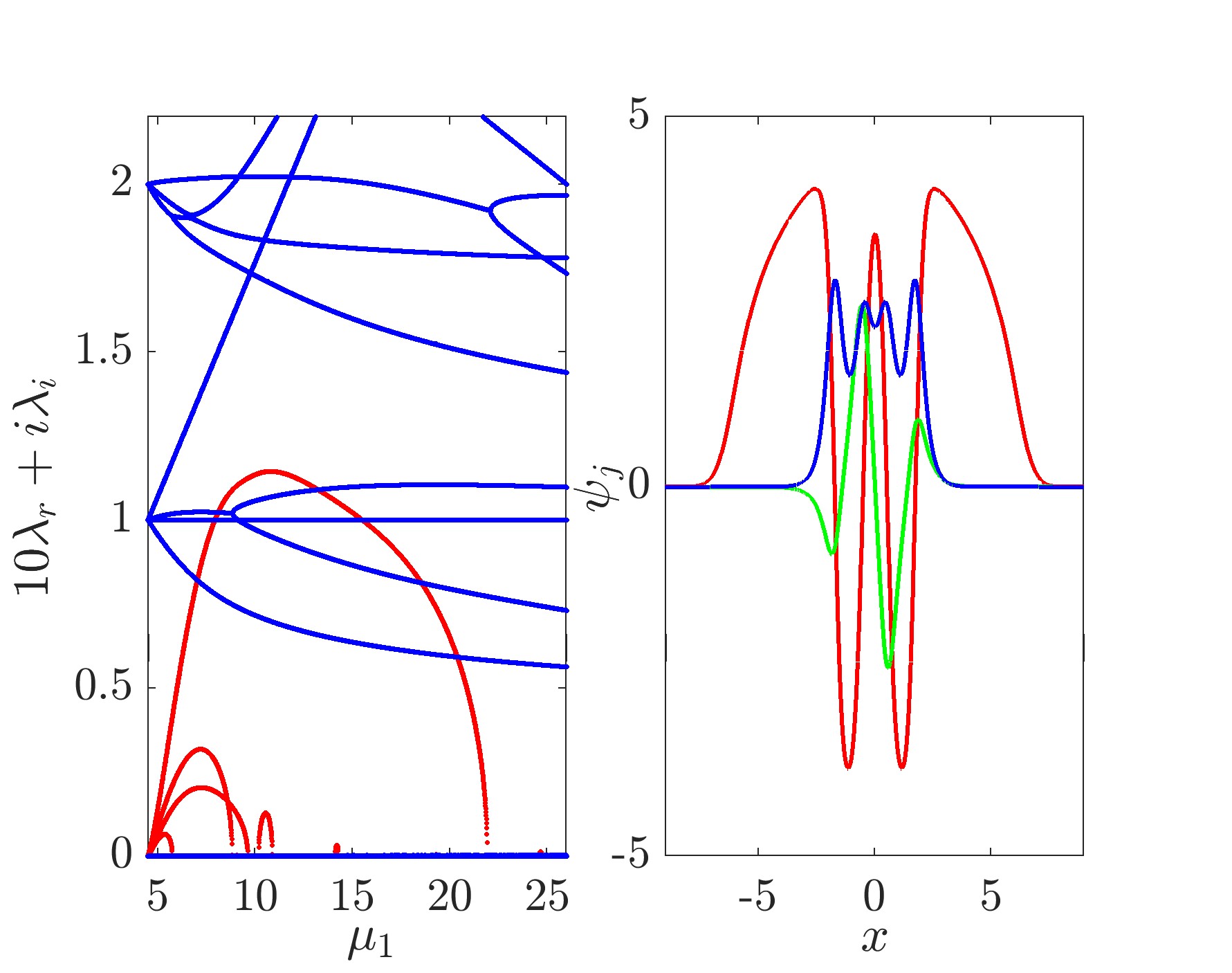}}
\subfigure[]{\includegraphics[width=0.33\textwidth]{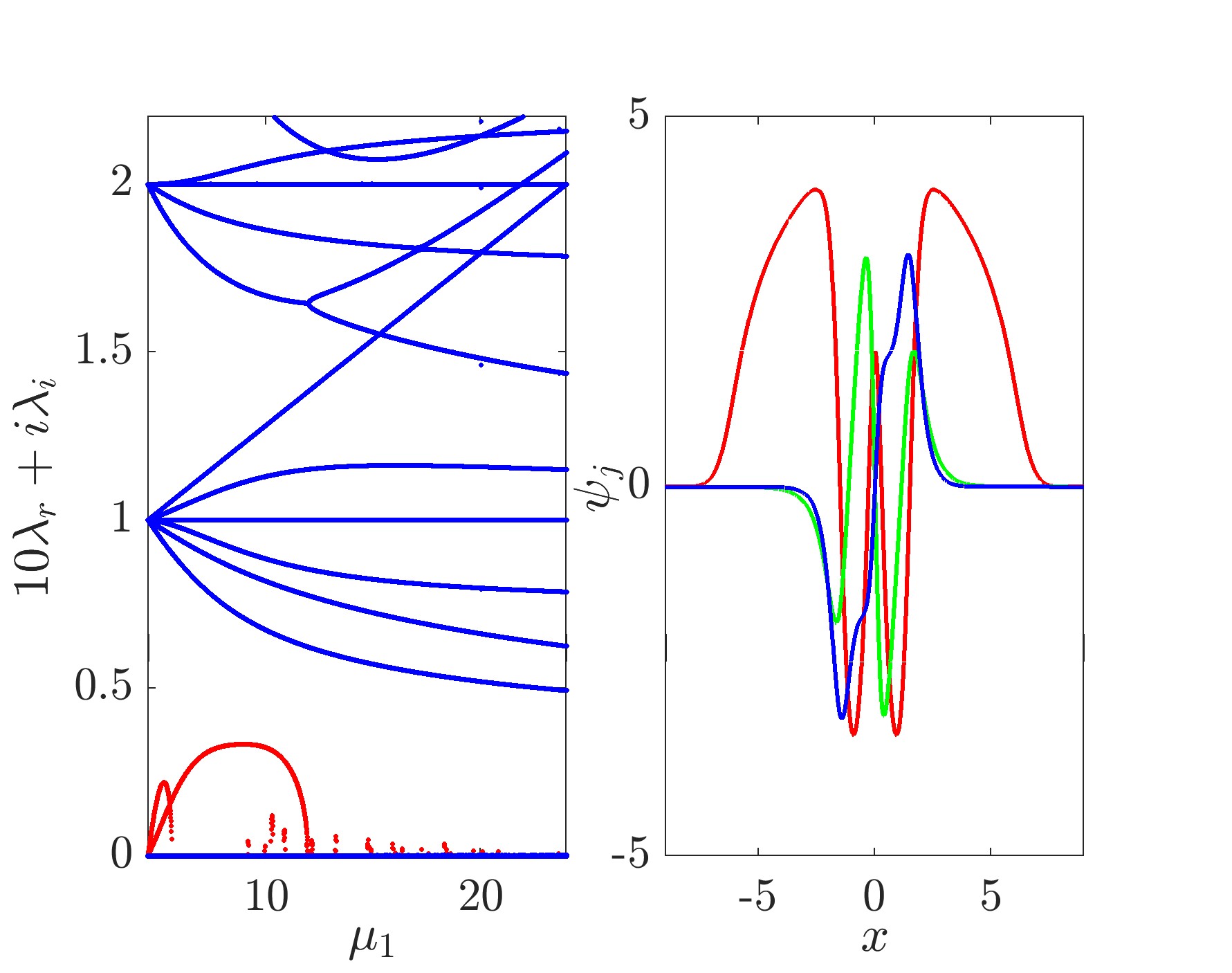}}
\subfigure[]{\includegraphics[width=0.33\textwidth]{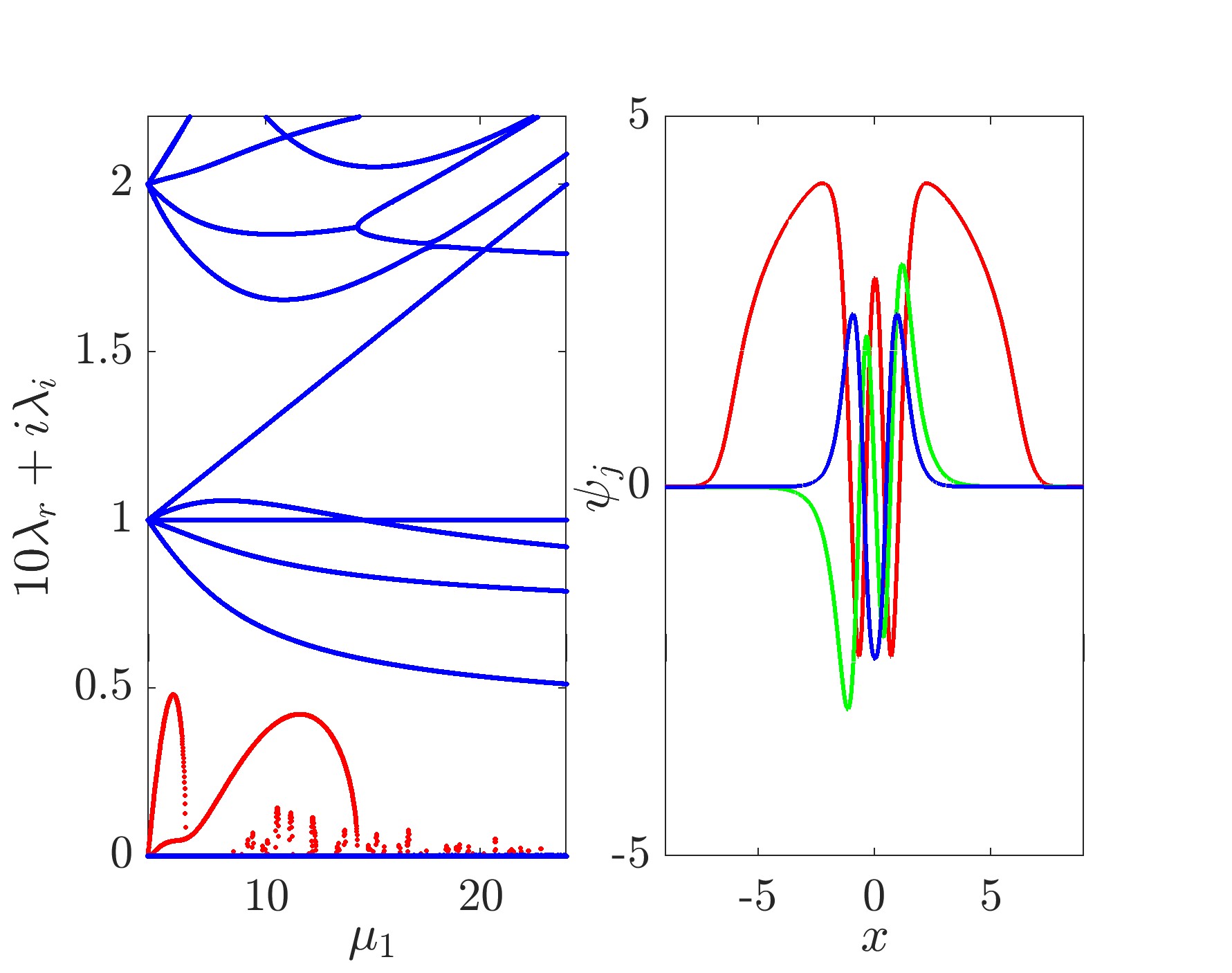}}
\caption{
Same as Fig.~\ref{S210} but for the $n_1=4$ family of states \state{410}, \state{420}, \state{421}, \state{430}, \state{431}, and \state{432}, respectively. The final chemical potentials are $(24,22,20)$, except for the \state{430} state. The state does not become fully stable along this trajectory, but can be stabilized when we extend it further into the Thomas-Fermi regime. Here, its final chemical potentials are only slightly larger $(26,22,20)$.
}
\label{S410}
\end{figure*}

\begin{figure*}
\subfigure[]{\includegraphics[width=0.24\textwidth]{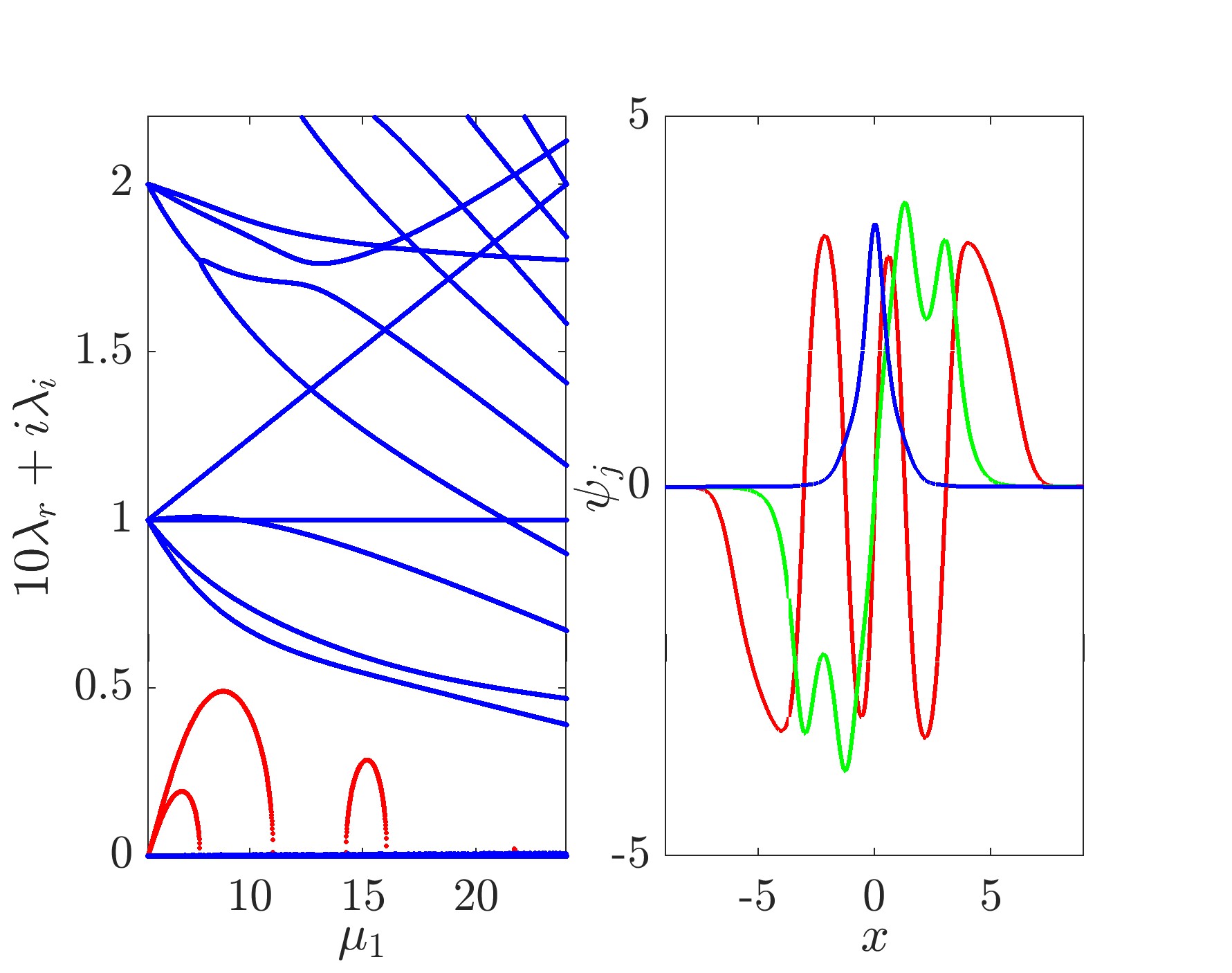}}
\subfigure[]{\includegraphics[width=0.24\textwidth]{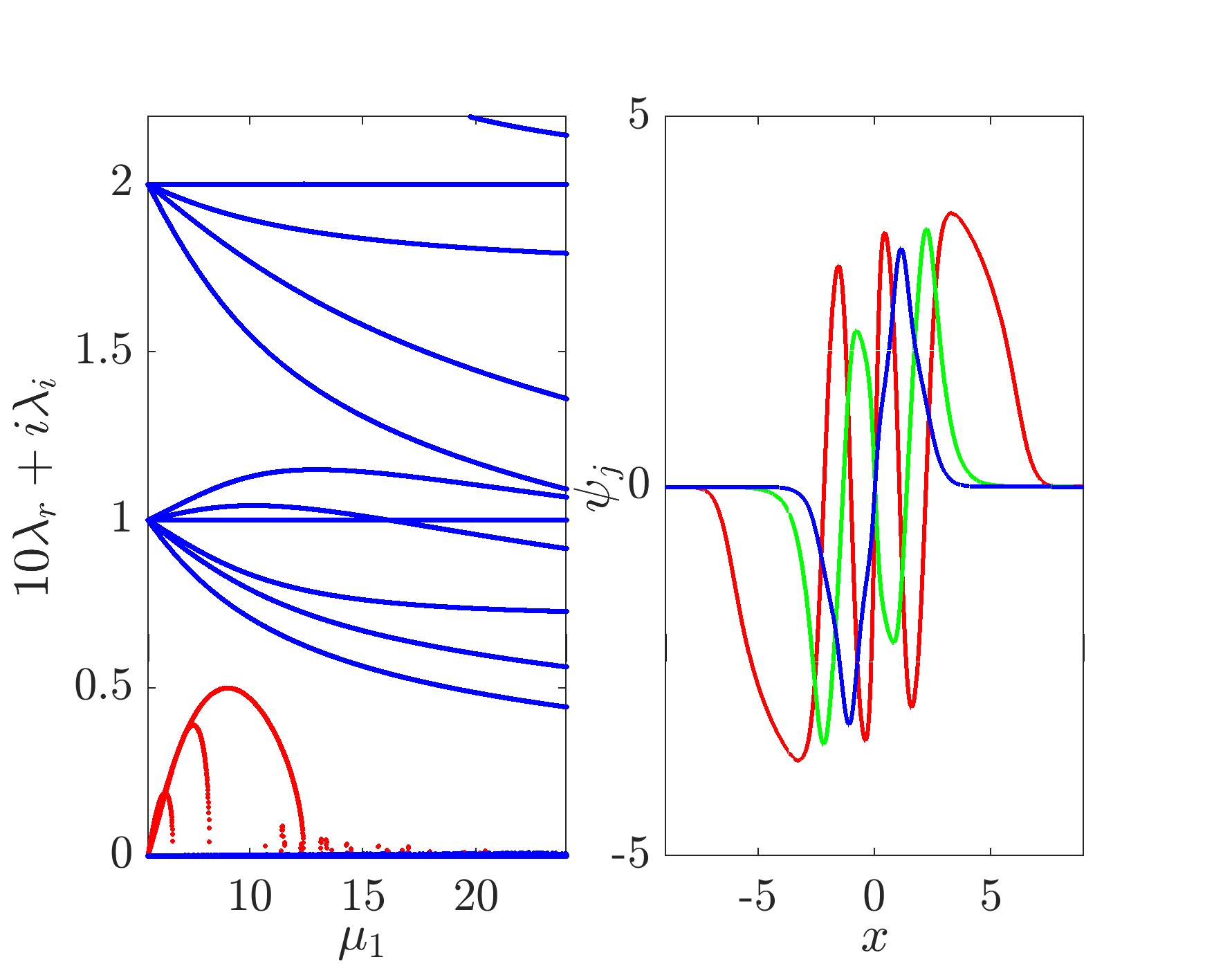}}
\subfigure[]{\includegraphics[width=0.24\textwidth]{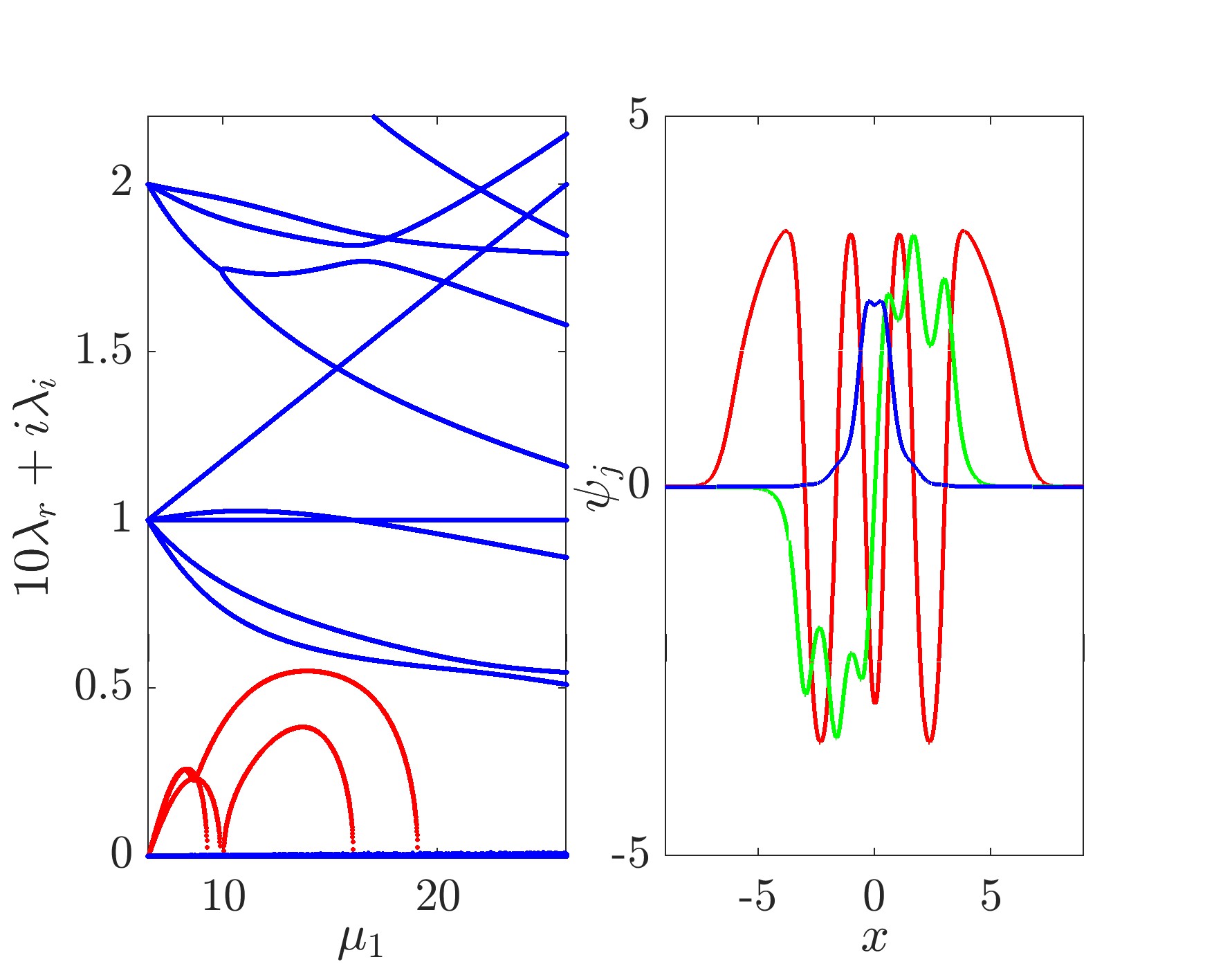}}
\subfigure[]{\includegraphics[width=0.24\textwidth]{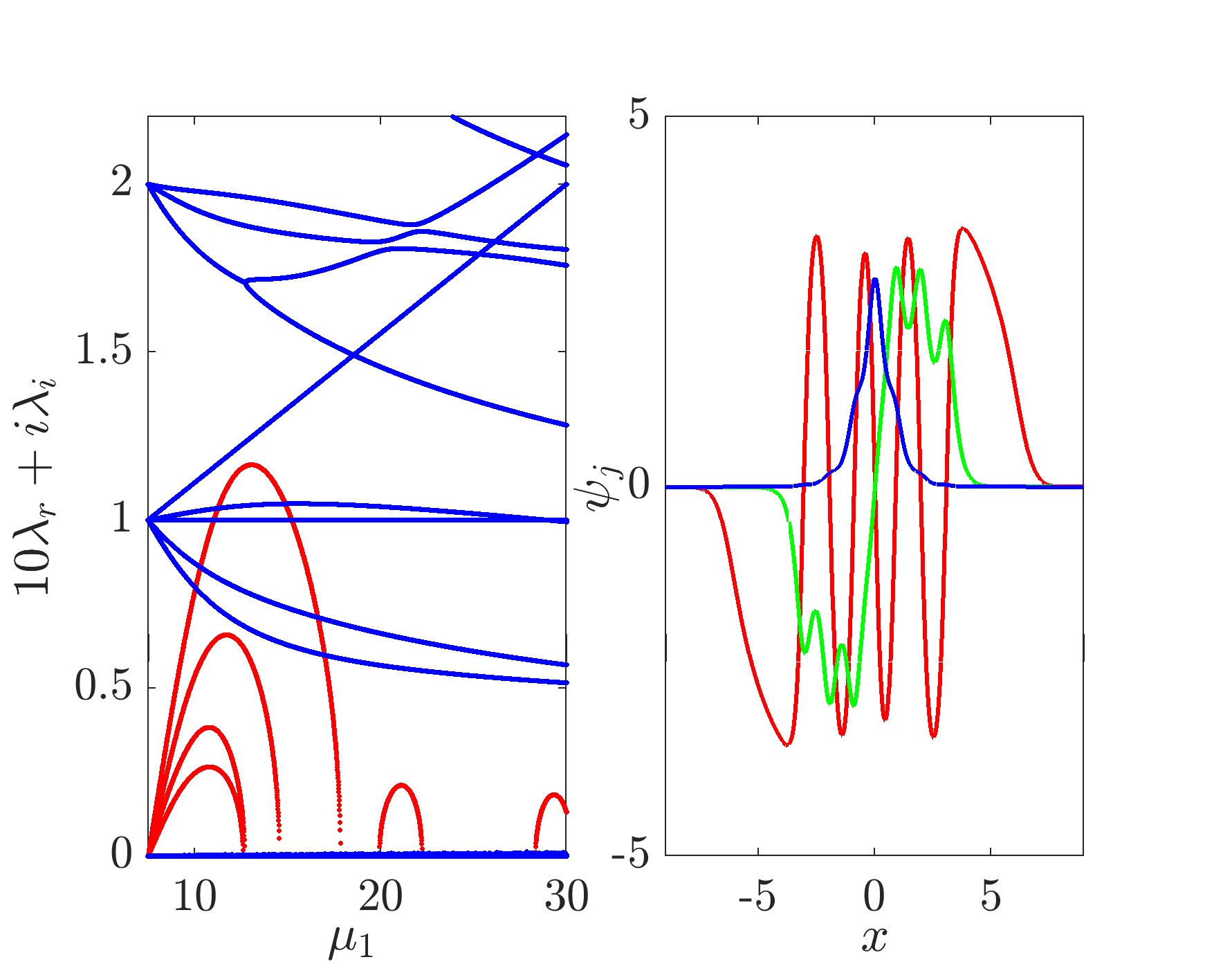}}
\caption{
Same as Fig.~\ref{S210} but for a few higher-lying states \state{510}, \state{531}, \state{610}, \state{710}, with final chemical potentials $(24,22,20)$, $(24,22,20)$, $(26,22,20)$, $(30,24,22)$, respectively. Note that the state \state{531} has all odd quantum numbers, and the center is a dark-dark-dark structure.
}
\label{S510}
\end{figure*}

\begin{figure*}
\subfigure[]{\includegraphics[width=0.33\textwidth]{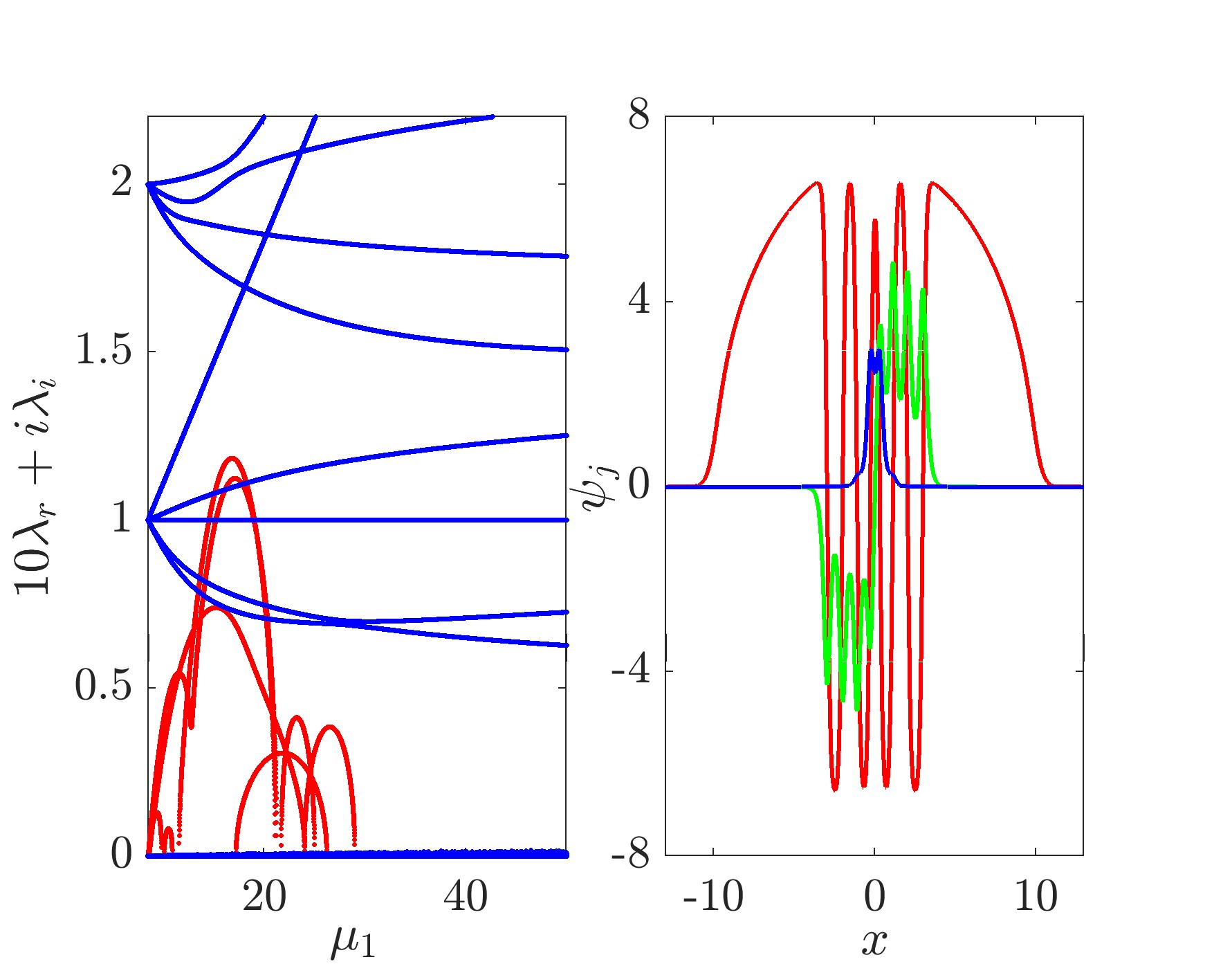}}
\subfigure[]{\includegraphics[width=0.33\textwidth]{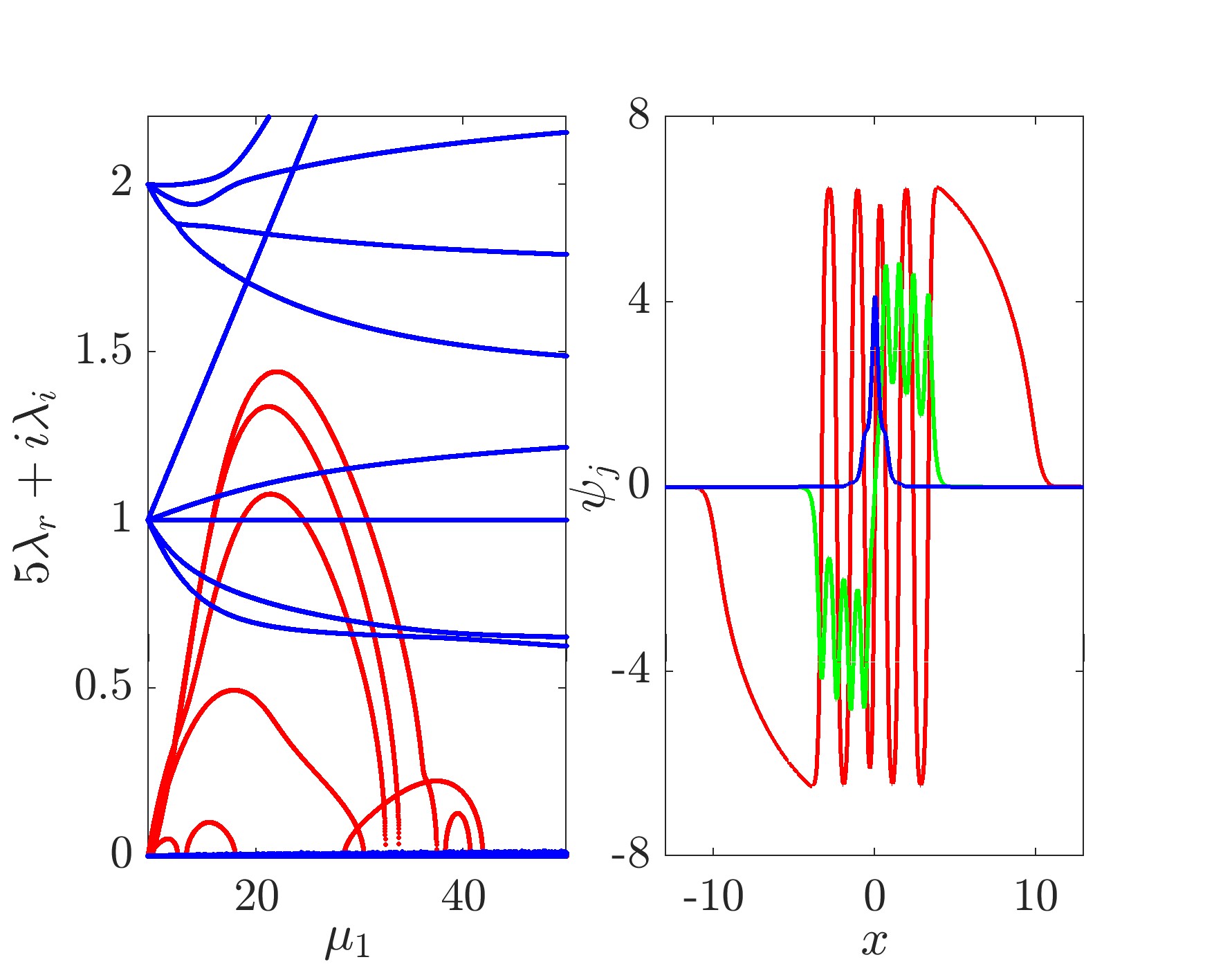}}
\subfigure[]{\includegraphics[width=0.33\textwidth]{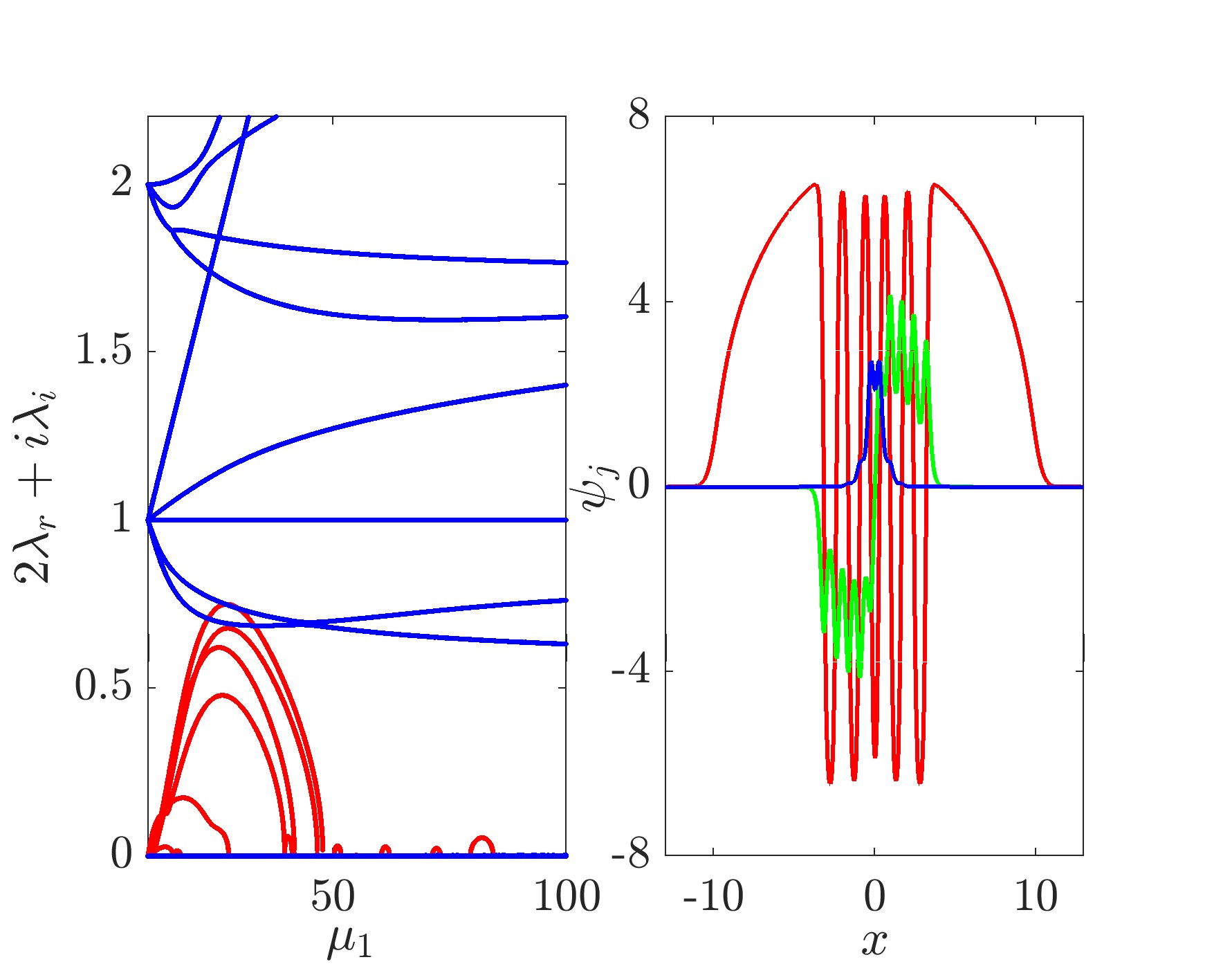}}
\caption{
Same as Fig.~\ref{S210} but for three highly excited states \state{810}, \state{910}, \state{1010}, with final chemical potentials $(50,36,32)$, $(50,36,32)$, $(100,70,64)$, respectively. Here, the depicted configurations are at $\mu_1$=50 instead. Note the different scaling factors for the real part of the eigenvalues. It is interesting that the third component is mostly concentrated in the center, while it is the second component that fills the off-center density dips of the dark soliton lattice, rendering the ``lattice'' rather heterogeneous.
}
\label{S810}
\end{figure*}

Next, there are a total of $6$ states in the family $n_1=4$, \state{410}, \state{420}, \state{421}, \state{430}, \state{431}, and \state{432}, as illustrated in Fig.~\ref{S410}. The former $2$ states are relatively robust, while the latter $4$ states are more prone to instabilities. Nevertheless, all of these states can be suitably stabilized. Similarly, the total density profile (not shown) again varies among the structures, and local density minimums are found and they are correlated with the density peaks of the trapped second and third components.

It is clearly impossible to exhaust all of the states, there is an infinity number of them, and thus we shall turn to representative ones in the following. We emphasize that this is only due to the large number of states available, and in this work we have not encountered any linear state that cannot be continued to its corresponding nonlinear counterpart. It seems that for a given $n_1$ the lowest-lying structure, i.e., \state{n_110}, has the best stability in the $n_1$ family. In the following, we focus on such states for simplicity. In addition, we also consider \state{531} which has a central dark-dark-dark structure, it is the lowest-lying state where all the quantum numbers are odd. These states are summarized in Figs.~\ref{S510} and~\ref{S810}. These states too can be fully stabilized, and we shall not repeatedly mention this fact every time as this is true for \textit{all} the structures we studied.

We can readily identify the well-known building blocks of the localized dark-dark-dark, dark-dark-bright, and dark-bright-bright structures \cite{engels18,Lichen:DT} in the obtained states. The central part of \state{531} is a dark-dark-dark soliton, while the other two structures are rather common. For example the central part of \state{210} is a dark-bright-bright soliton and the central part of \state{310} is a dark-dark-bright soliton, and such structures are also prevalent off the center. It is perhaps even more interesting that the decomposition of a solitary wave into an array of localized structures is, however, frequently not relevant, contrary to one- and two-component structures. One striking example is the state \state{1010}, the third component is highly localized and the ``lattice'' is consequently rather heterogeneous. The second component fills the density dips of the dark soliton lattice at the sides while the third component fills the density dips around the center, the third component essentially disappears at the sides of the lattice, rendering the sides effectively two-component dark-bright lattices.

As the states get increasingly complex, they also become harder to stabilize, which appears to be a generic feature \cite{Wang:DD}. There are more unstable modes and the growth rates are also larger. The state \state{1010} becomes relatively robust only when $\mu_1 \gtrsim 50$. As the number of components grows, the spectrum is also more sensitive to the particular choice of the final chemical potentials or the continuation path, compared with one- and two-component systems. We find that a structure tends to be more readily stabilized if the peak densities, which are correlated with the respective chemical potentials, do not vary significantly between adjacent components.

\begin{figure*}
\subfigure[]{\includegraphics[width=0.33\textwidth]{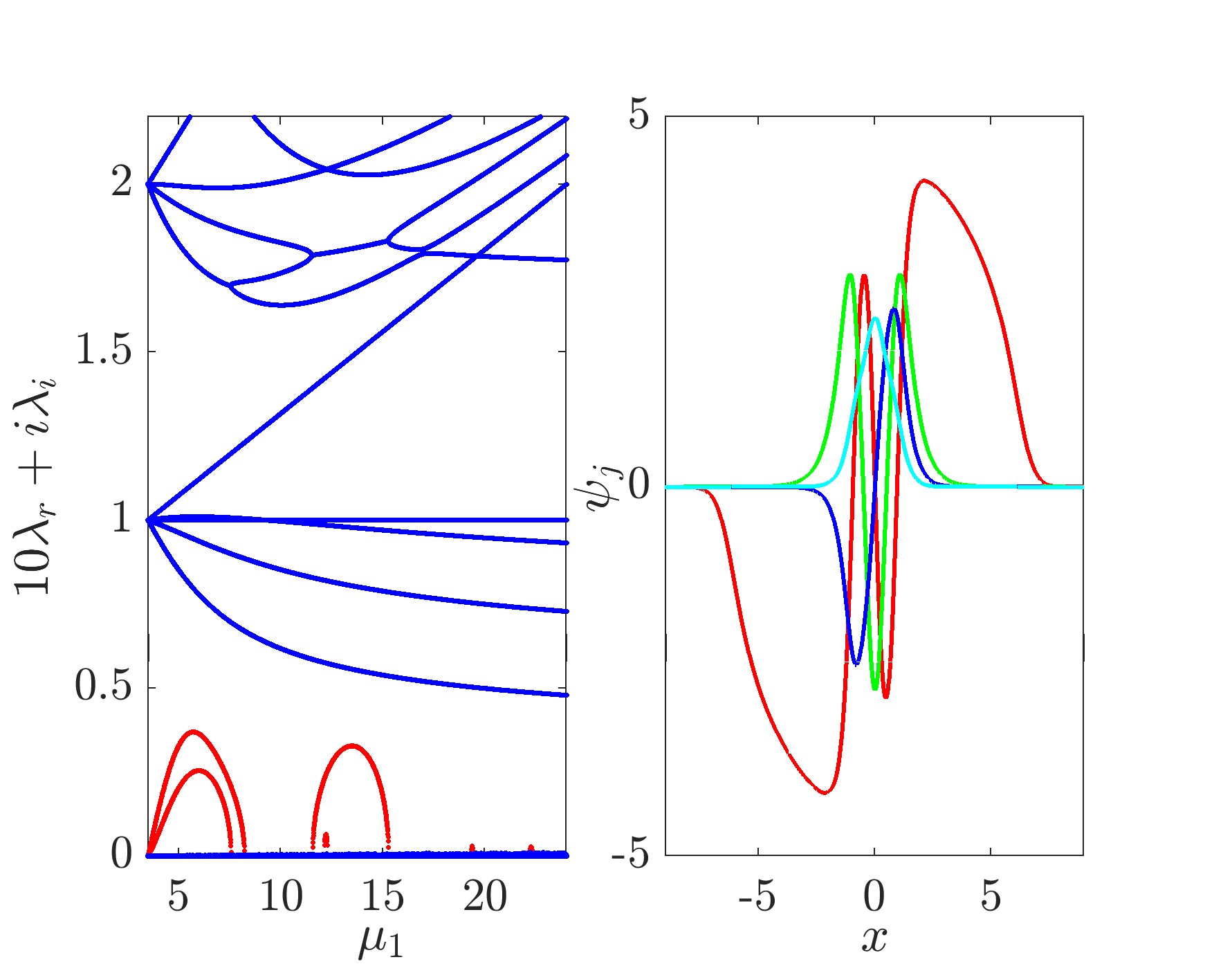}}
\subfigure[]{\includegraphics[width=0.33\textwidth]{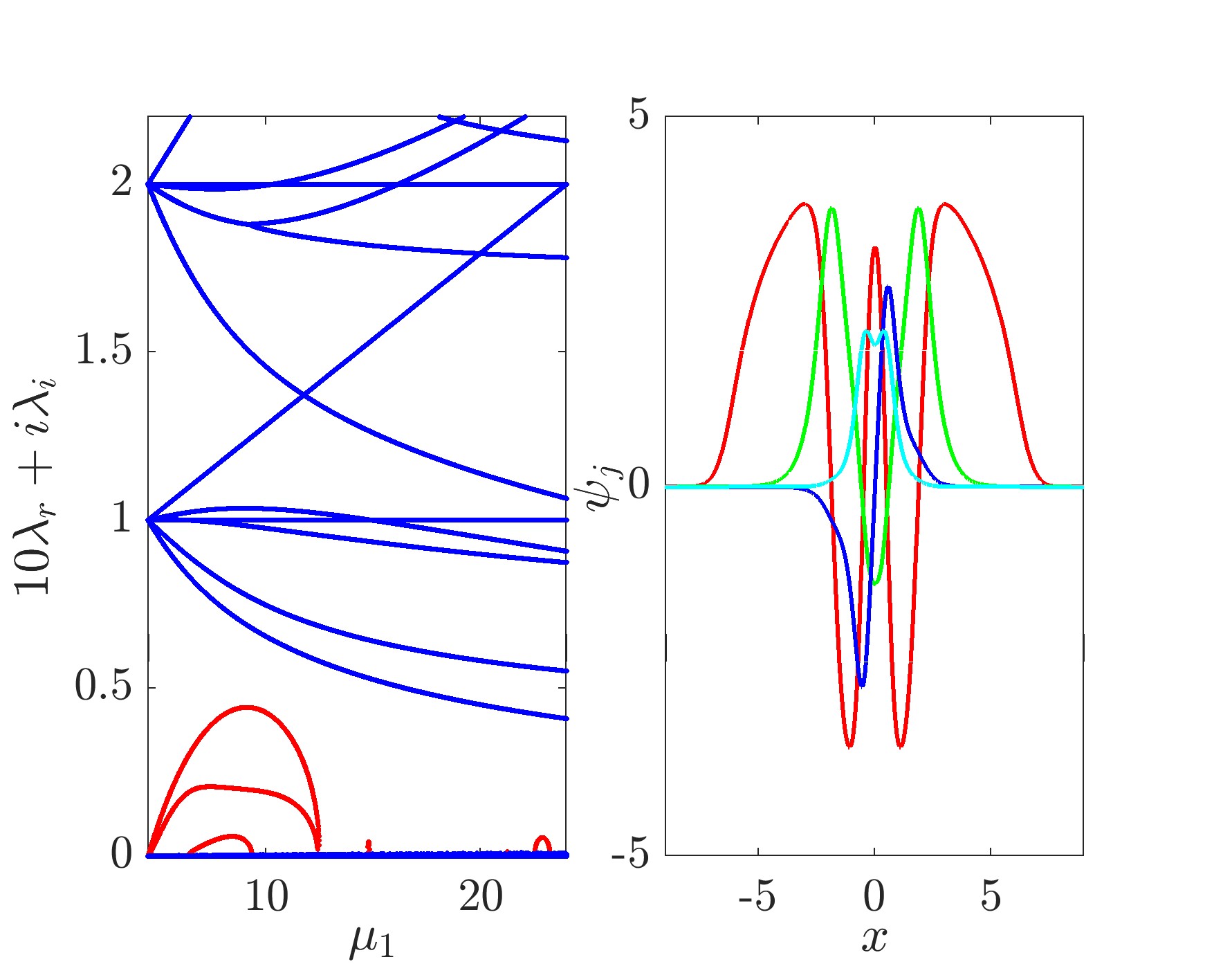}}
\subfigure[]{\includegraphics[width=0.33\textwidth]{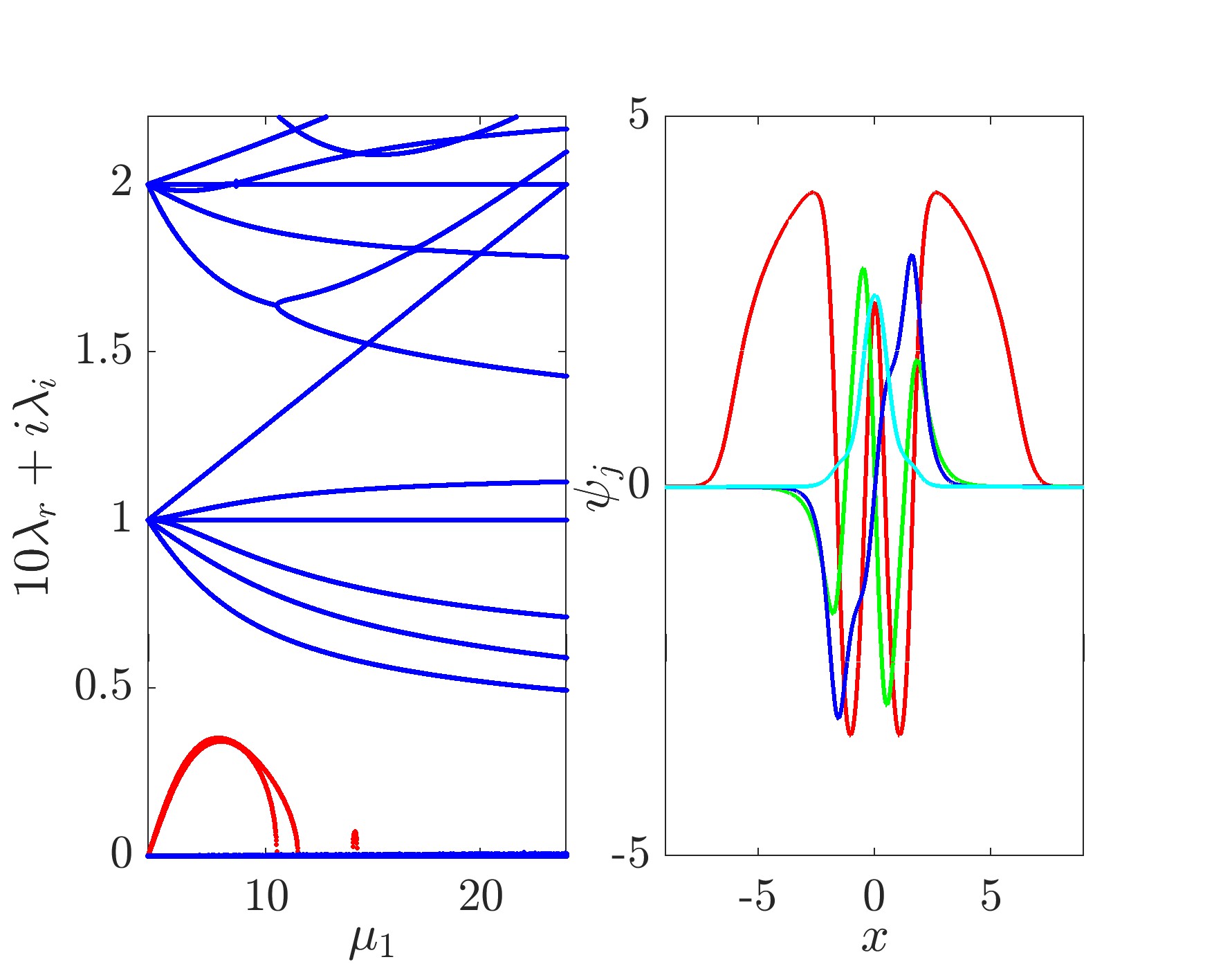}}
\subfigure[]{\includegraphics[width=0.33\textwidth]{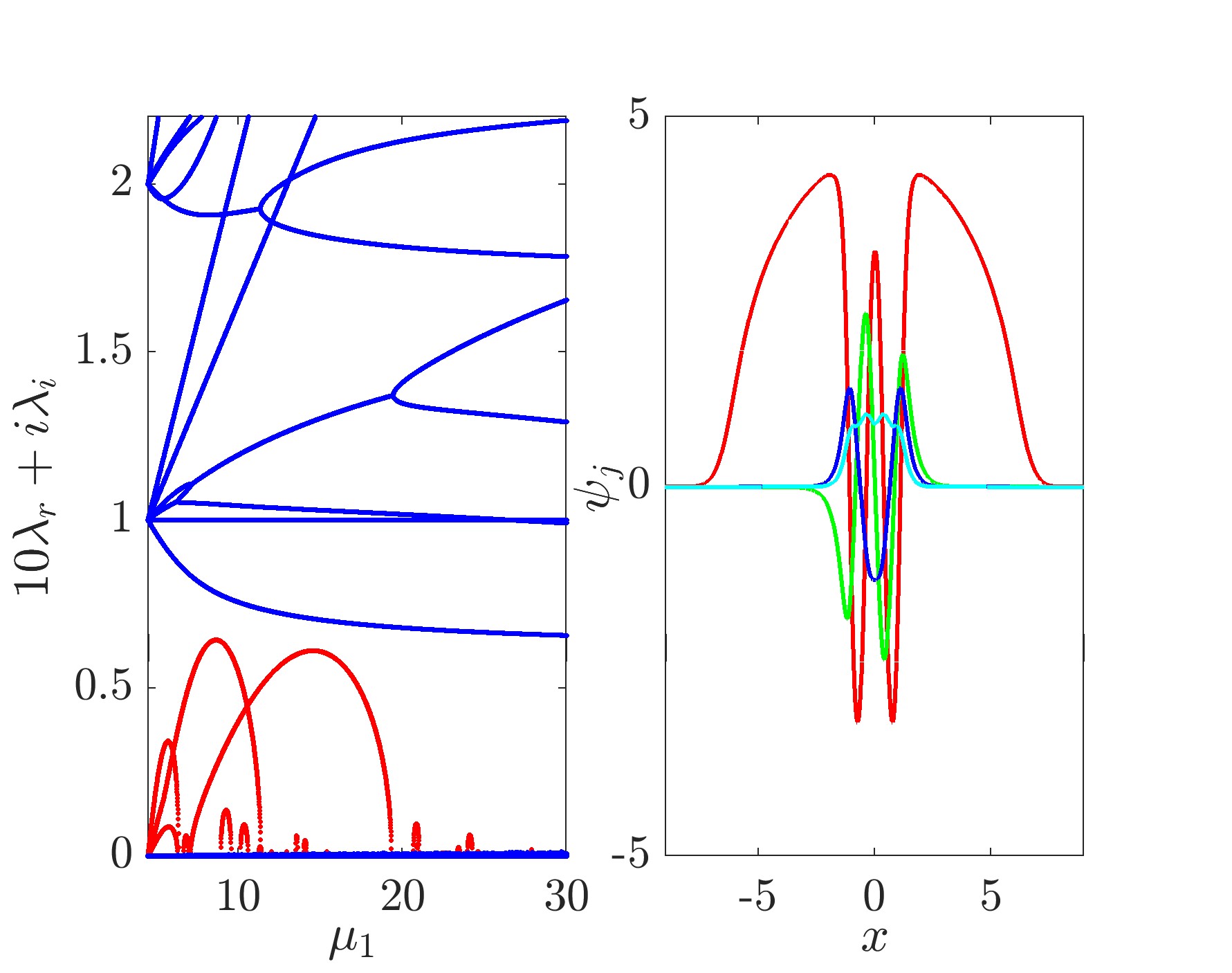}}
\subfigure[]{\includegraphics[width=0.33\textwidth]{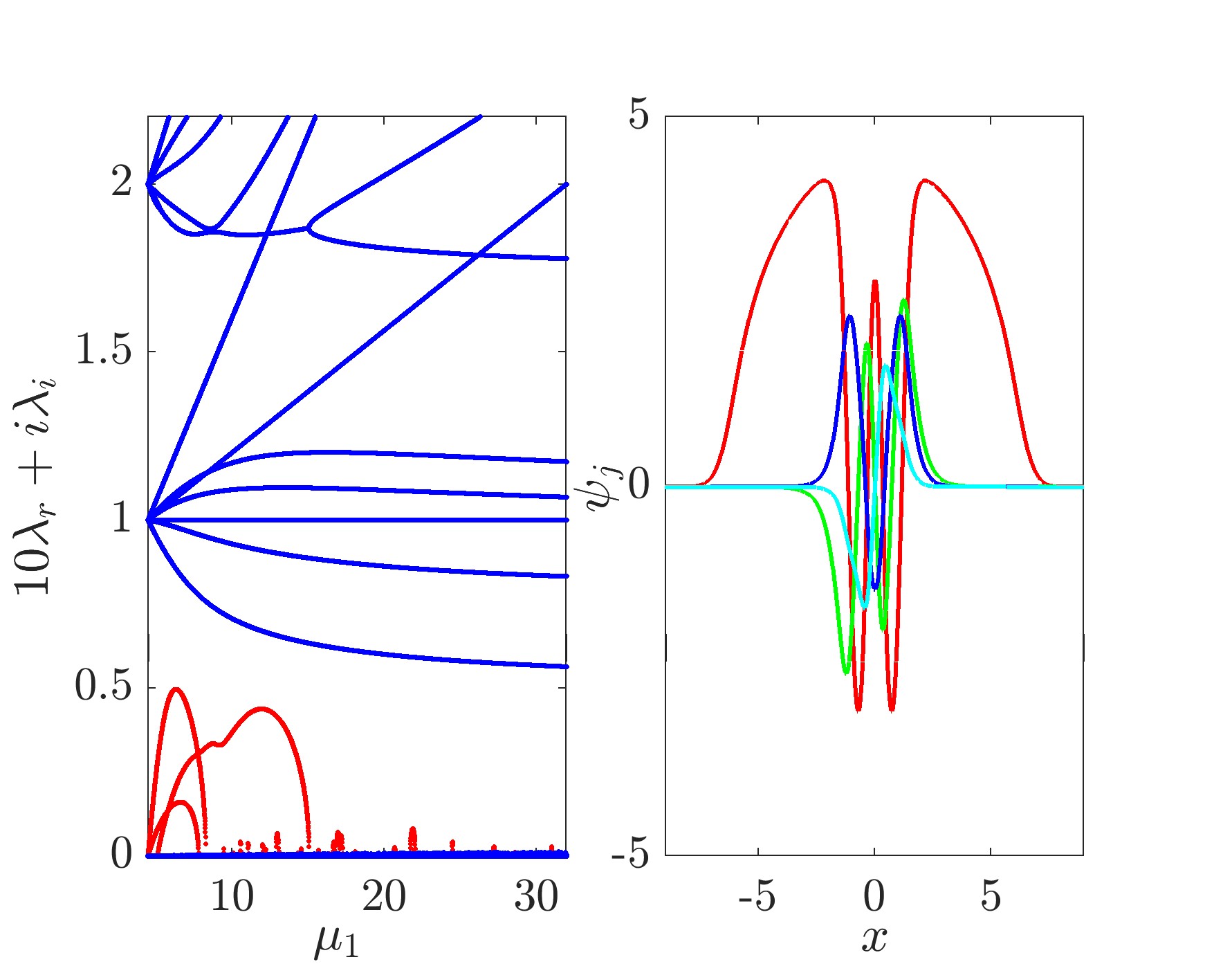}}
\subfigure[]{\includegraphics[width=0.33\textwidth]{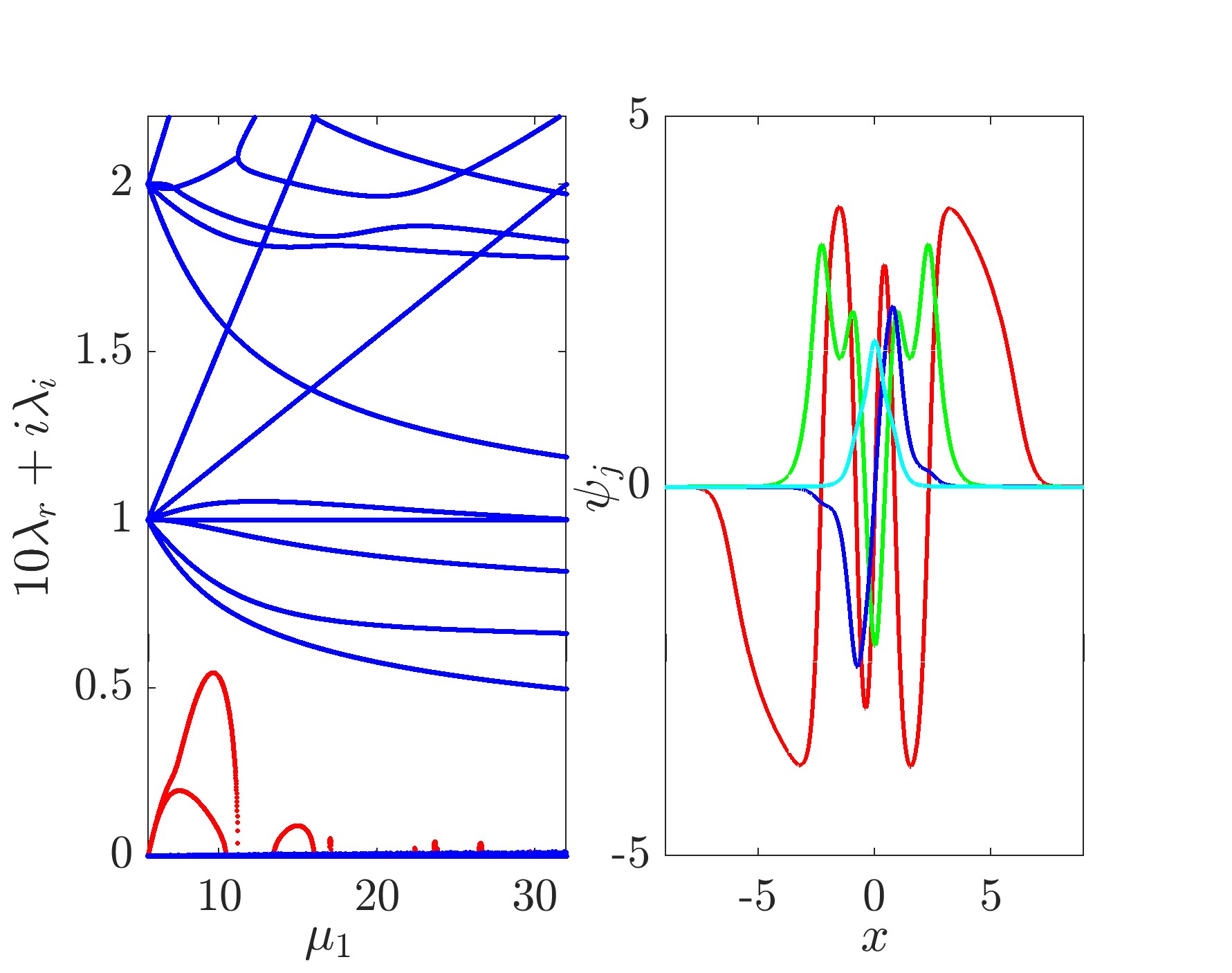}}
\caption{
The first $6$ low-lying four-component states \state{3210}, \state{4210}, \state{4310}, \state{4320}, \state{4321}, \state{5210} with final chemical potentials $(24,22,20,18)$, $(24,22,20,18)$, $(24,22,20,18)$, $(30,24,20,16)$, $(32,28,26,22)$ , $(32,28,26,22)$, respectively. The depicted states are at $\mu_1=20$, and the fourth component is shown in cyan.
}
\label{S3210}
\end{figure*}

\begin{figure*}
\subfigure[]{\includegraphics[width=0.33\textwidth]{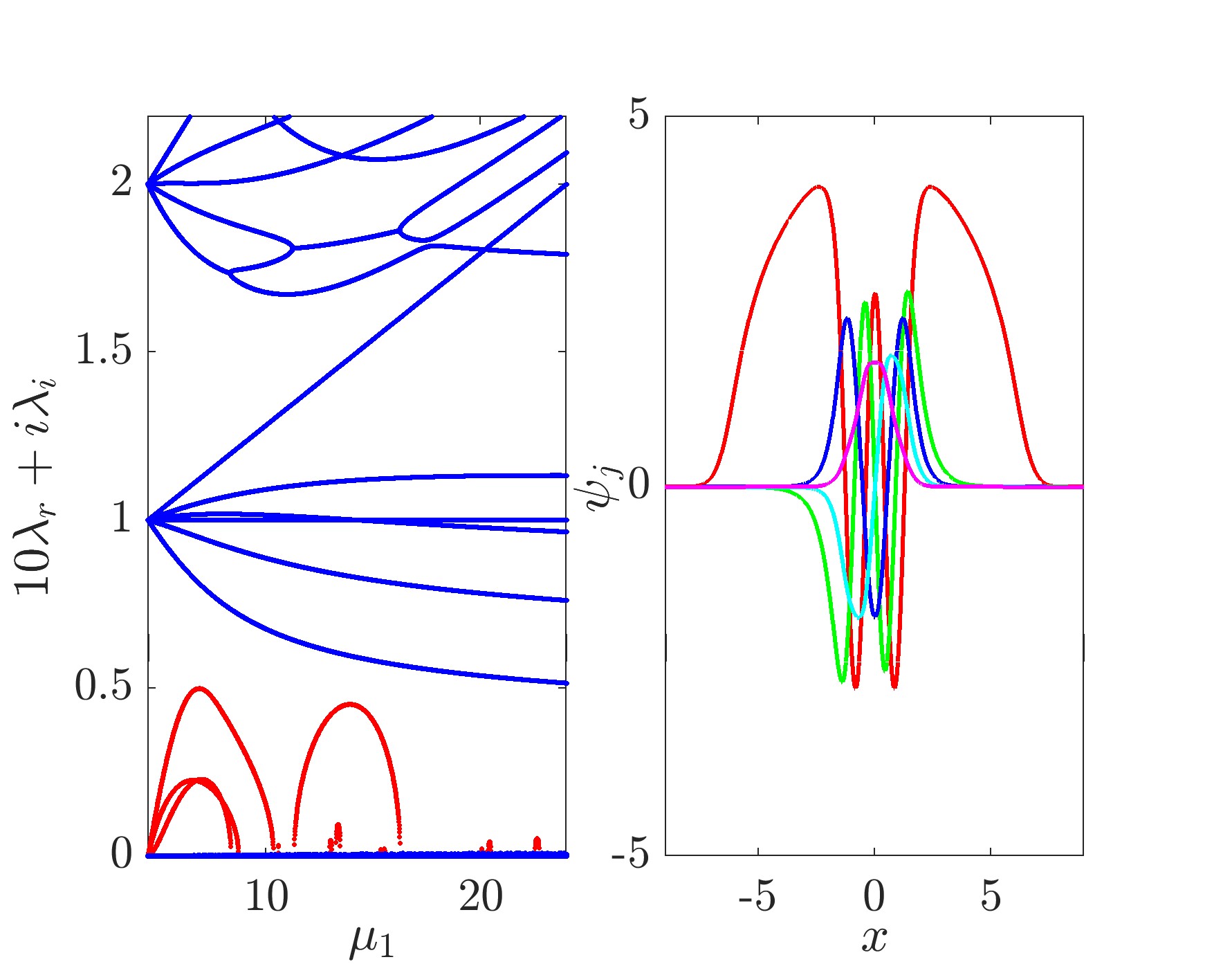}}
\subfigure[]{\includegraphics[width=0.33\textwidth]{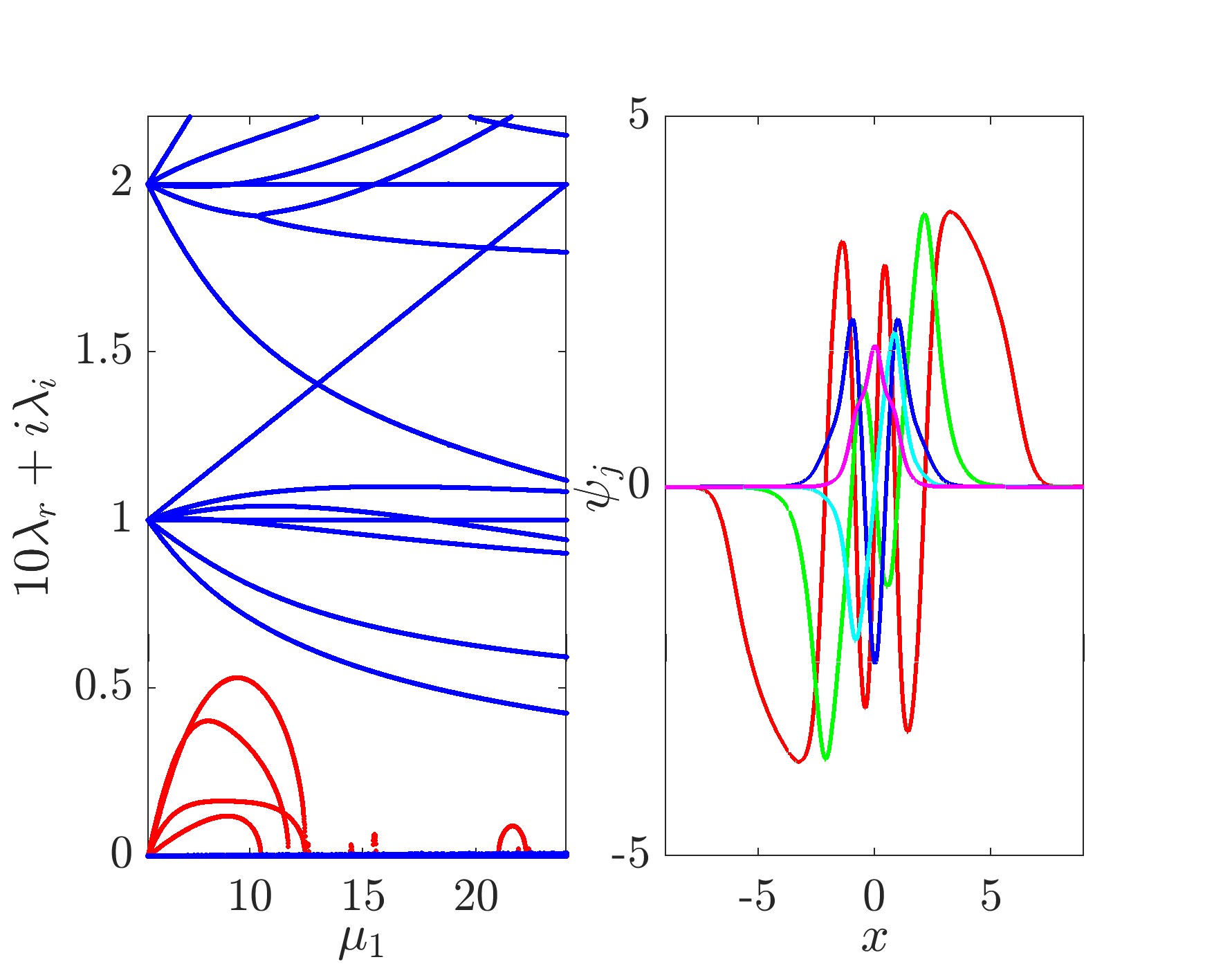}}
\subfigure[]{\includegraphics[width=0.33\textwidth]{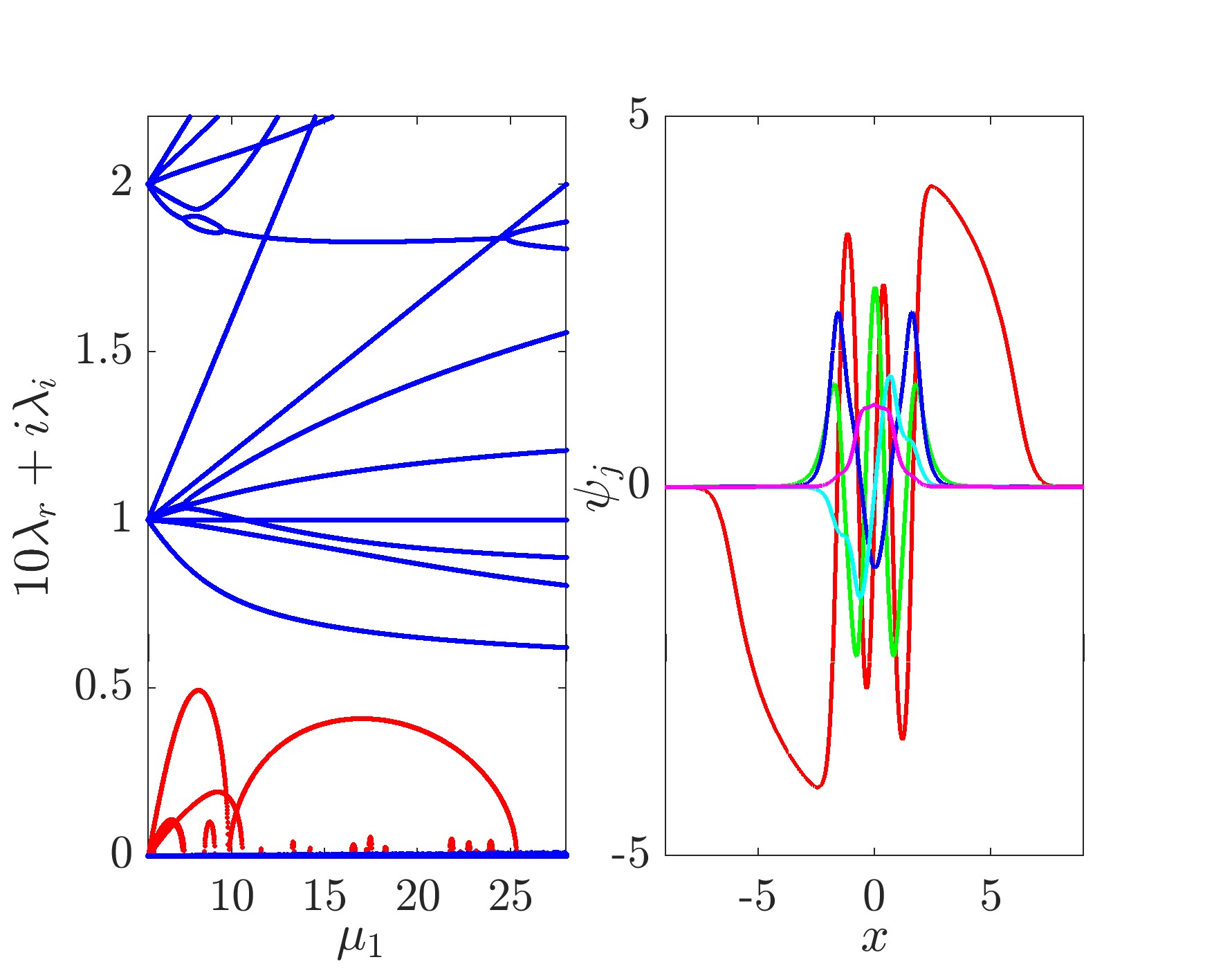}}
\subfigure[]{\includegraphics[width=0.33\textwidth]{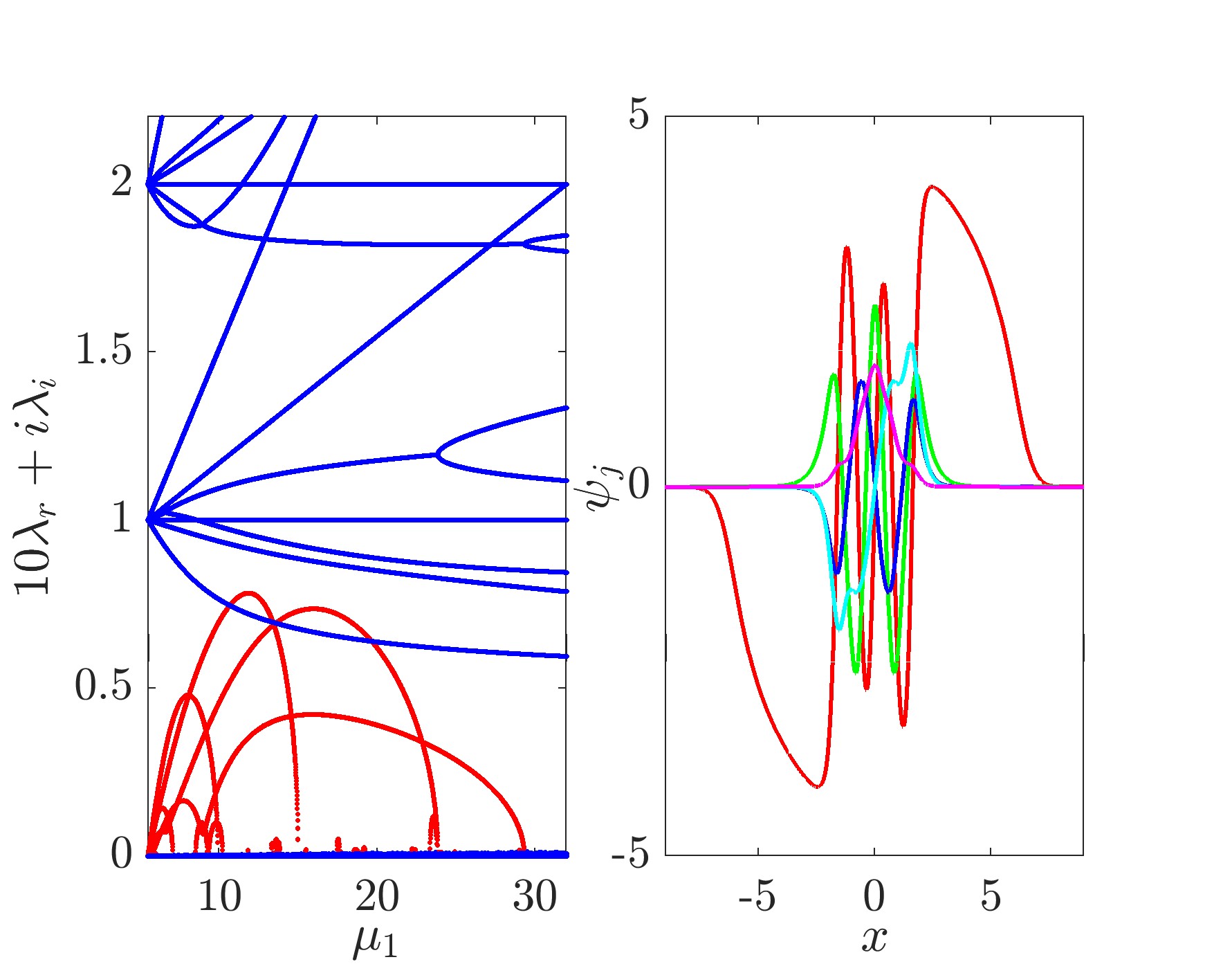}}
\subfigure[]{\includegraphics[width=0.33\textwidth]{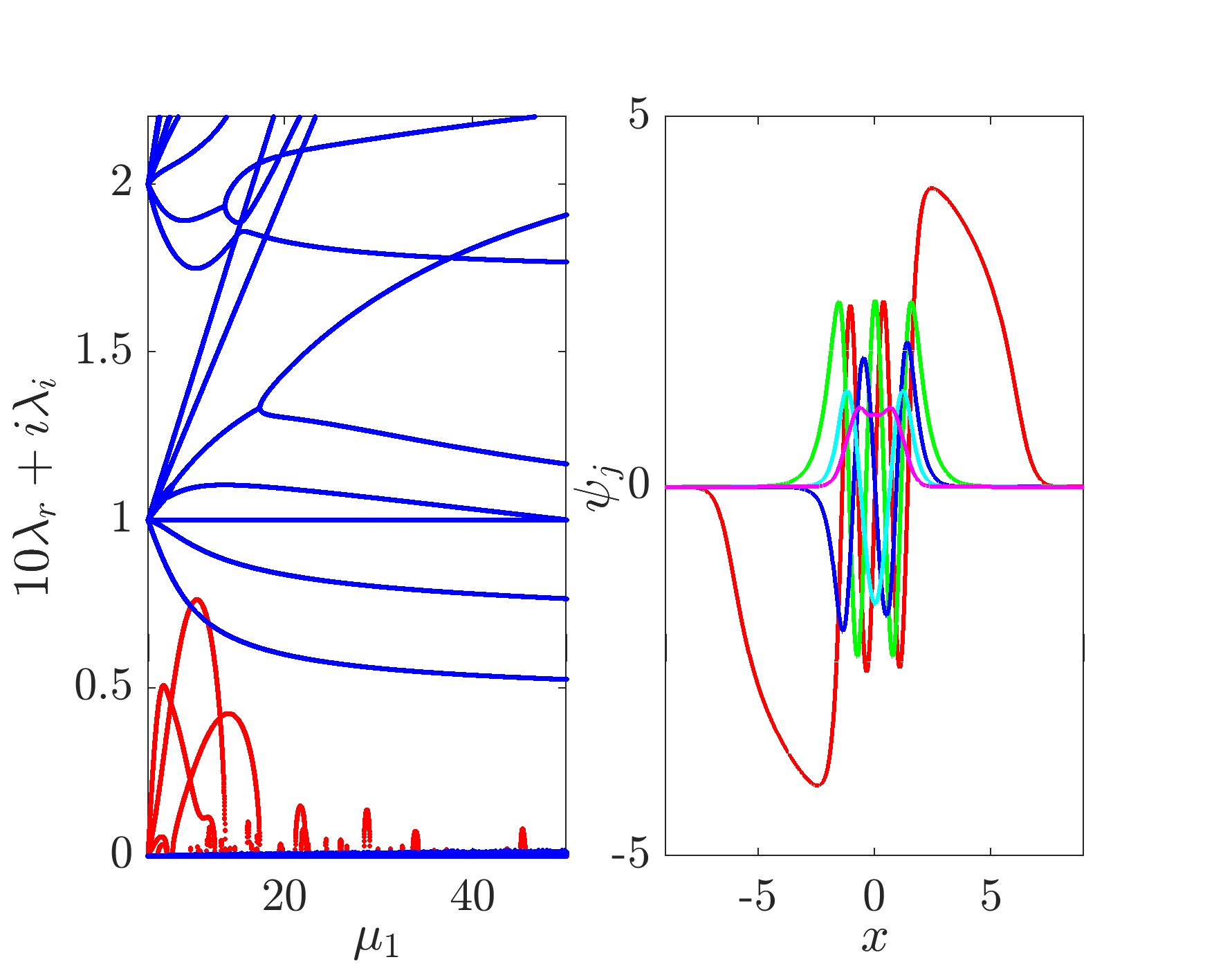}}
\subfigure[]{\includegraphics[width=0.33\textwidth]{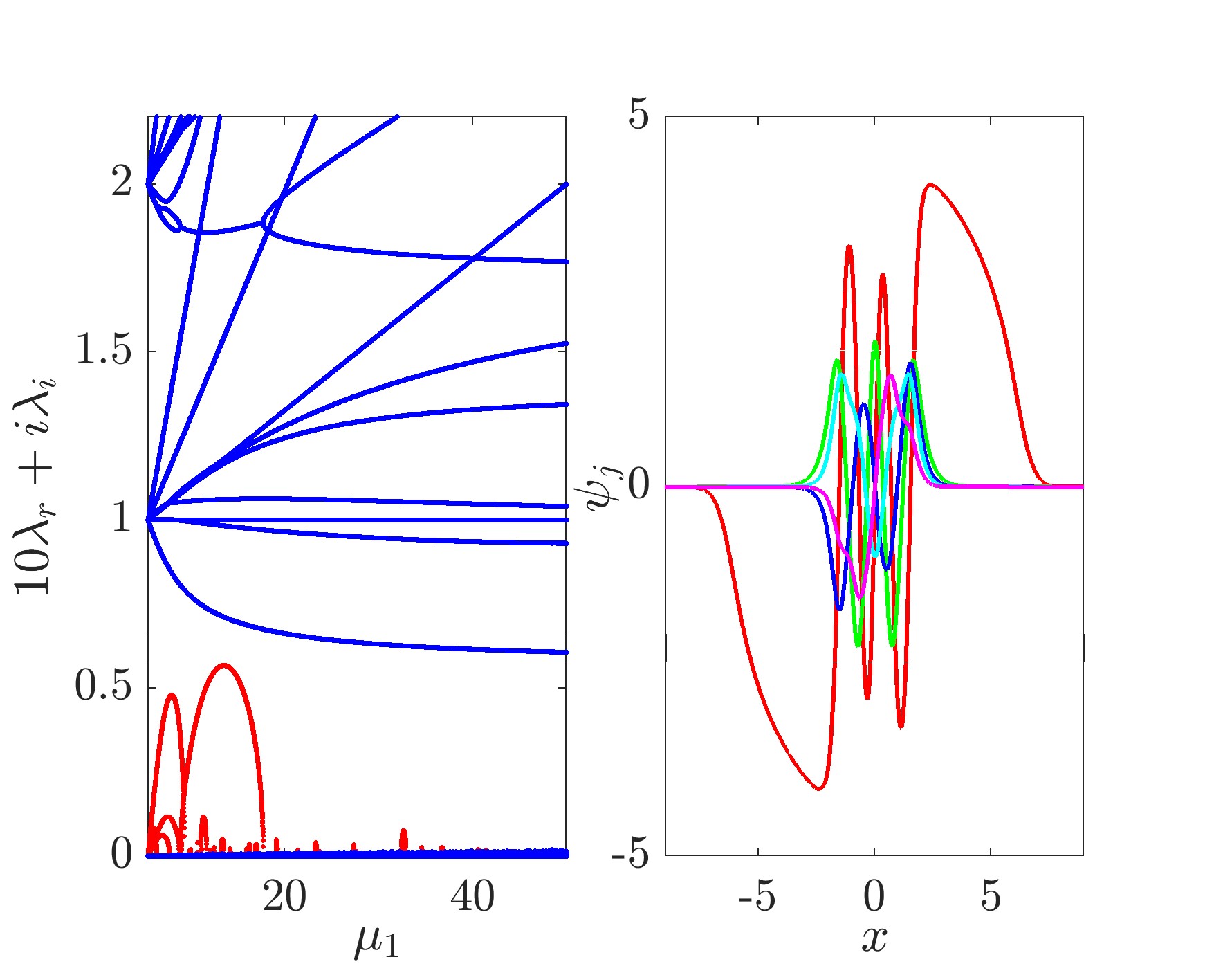}}
\caption{
The first $6$ low-lying five-component states \state{43210}, \state{53210}, \state{54210}, \state{54310}, \state{54320}, \state{54321} with final chemical potentials $(24,22,20,18,16)$, $(24,22,20,18,16)$, $(28,24,20,18,16)$, $(32,28,24,22,20)$, $(50,46,41,36,30)$, $(50,42,38,36,34)$, respectively. The depicted states are at $\mu_1=20$, and the fifth component is shown in pink.
}
\label{S43210}
\end{figure*}

Finally, the linear limit continuation can be readily applied to four- and five-component systems, despite that the induced solitary waves inevitably become increasingly complex in their structures, and their numbers also grow much more rapidly, see Table~\ref{table}. Here, we present the $6$ lowest-lying states for each setting in Figs.~\ref{S3210} and~\ref{S43210}, respectively. While none of these structures is fully robust in general, which is not surprising as this is already the case for three components, it is nevertheless remarkable that they again can be fully stabilized in suitable chemical potential intervals.

\subsection{SU($n$)-induced and driving-induced dynamics}

It is clearly not the goal of this work to explore the detailed properties of each of the continued solitary waves. Indeed, the large array of solitary waves constructed here provides an ideal setting for further studies of their properties, which are important in their own rights. Here, we present a few proof-of-principle dynamics of these solitary waves. We focus on two types but more are possible \cite{revip}, one is the $SU(n)$-induced beating dynamics \cite{Yan:DD,Lichen:DD,Wang:DD} and the other is driving-induced dynamics \cite{Lichen:DB}. Both lead to oscillatory dark solitons, but they have quite different features and mechanisms. For the driving-induced dynamics, we apply a constant force for simplicity to the ``bright solitons'', i.e., to the component stemming from the $|0\rangle$ state. 
For the $SU(n)$-induced dynamics, we use again for simplicity the following relatively symmetric $SU(2)$ (for a subrotation) and $SU(n)$ rotations:
\begin{eqnarray}
U_{2\times2} = \frac{1}{\sqrt{2}}
\left(
\begin{array}{ccc}
1 & 1 \\
1 & -1
\end{array}
\right),
\end{eqnarray}
\begin{eqnarray}
U_{3\times3} = \frac{1}{\sqrt{3}}
\left(
\begin{array}{ccc}
1 & 1 & 1 \\
1 & w_1 & w_2 \\
1 & w_2 & w_1
\end{array}
\right),
\end{eqnarray}
\begin{eqnarray}
U_{4\times4} = \frac{1}{2}
\left(
\begin{array}{cccc}
1 & 1 & 1 & 1 \\
1 & 1 & -1 & -1 \\
1 & -1 & 1 & -1 \\
1 & -1 & -1 & 1
\end{array}
\right),
\end{eqnarray}
\begin{eqnarray}
U_{5\times5} = \frac{1}{\sqrt{5}}
\left(
\begin{array}{ccccc}
1 & 1 & 1 & 1 & 1 \\
1 & z_1 & z_2 & z_3 & z_4 \\
1 & z_2 & z_4 & z_1 & z_3 \\
1 & z_3 & z_1 & z_4 & z_2 \\
1 & z_4 & z_3 & z_2 & z_1
\end{array}
\right),
\end{eqnarray}
where  $w_k=\exp(i2k\pi/3)$, $k=1$, $2$, and $z_k=\exp(i2k\pi/5)$, $k=1$, $2$, $3$, $4$.

\begin{figure*}
\subfigure[]{\includegraphics[width=0.33\textwidth]{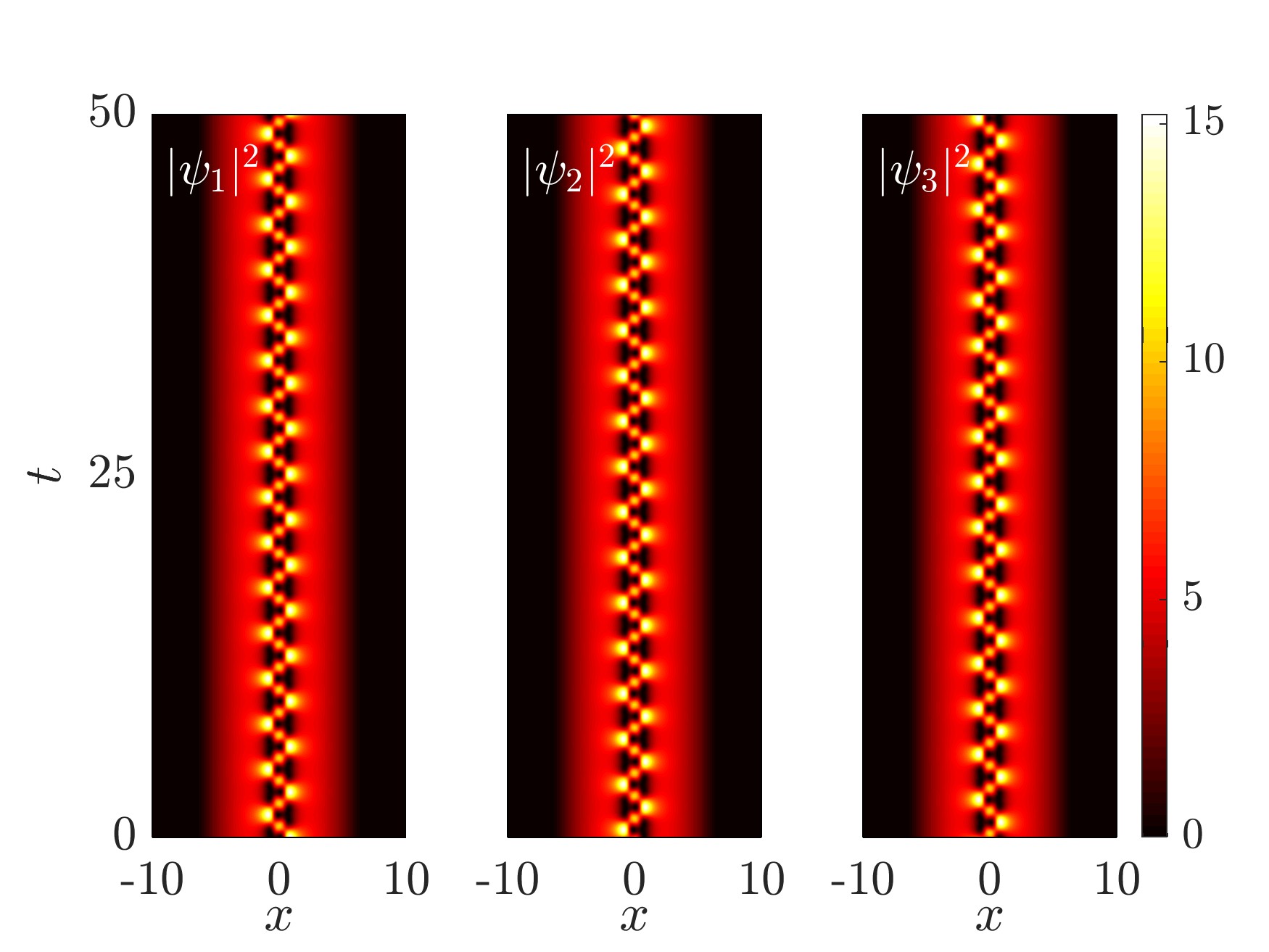}}
\subfigure[]{\includegraphics[width=0.33\textwidth]{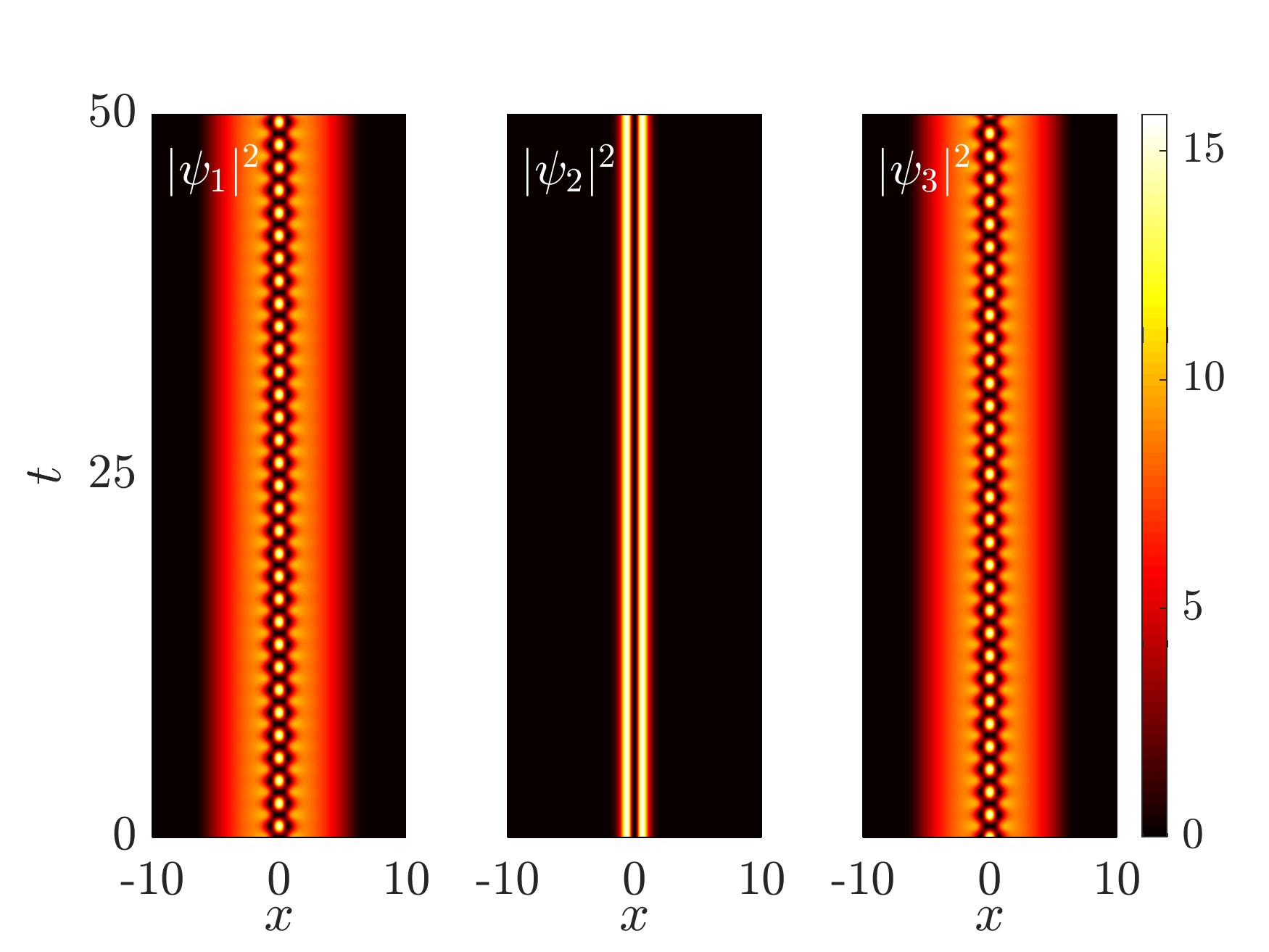}}
\subfigure[]{\includegraphics[width=0.33\textwidth]{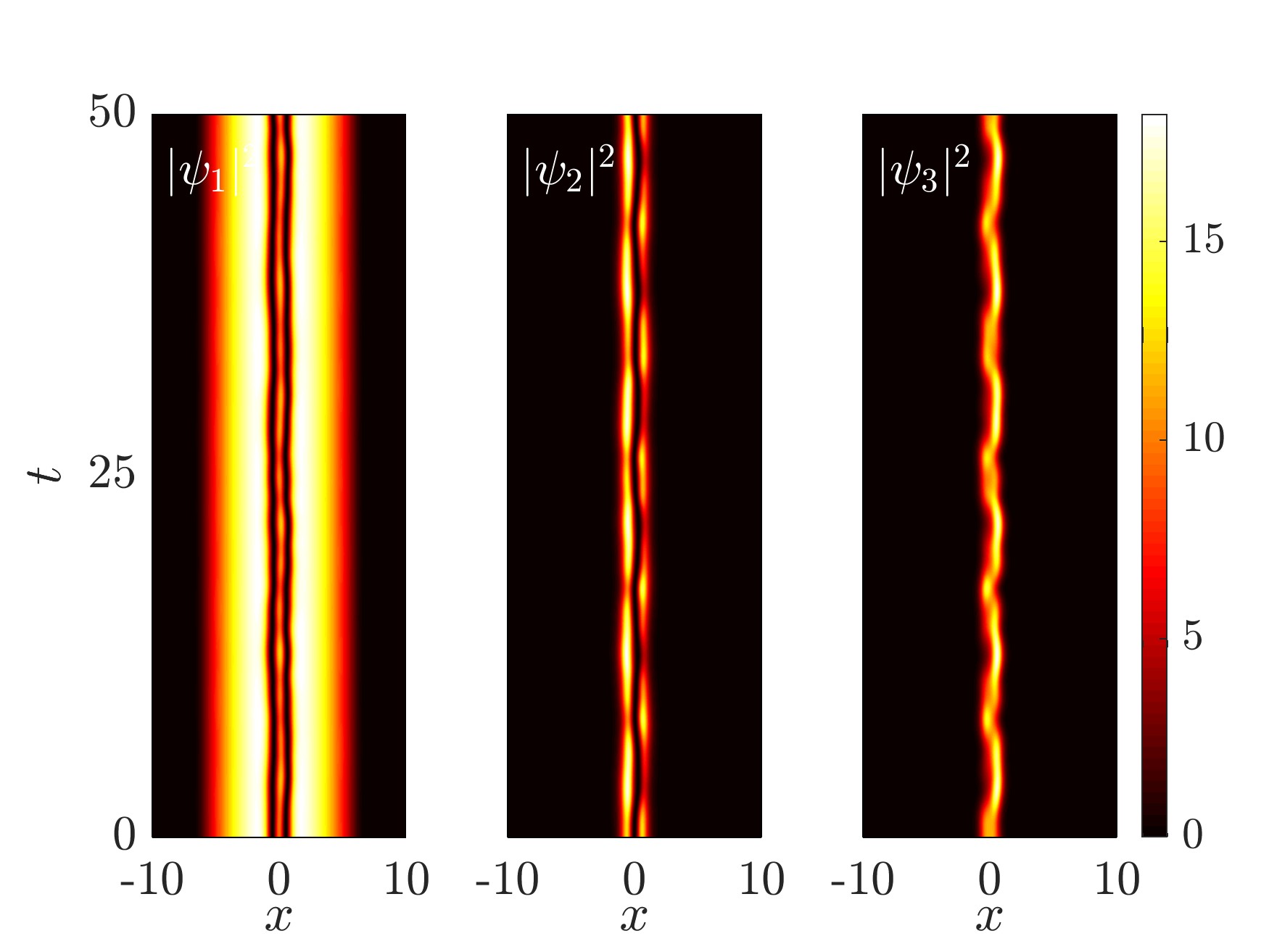}}
\subfigure[]{\includegraphics[width=0.33\textwidth]{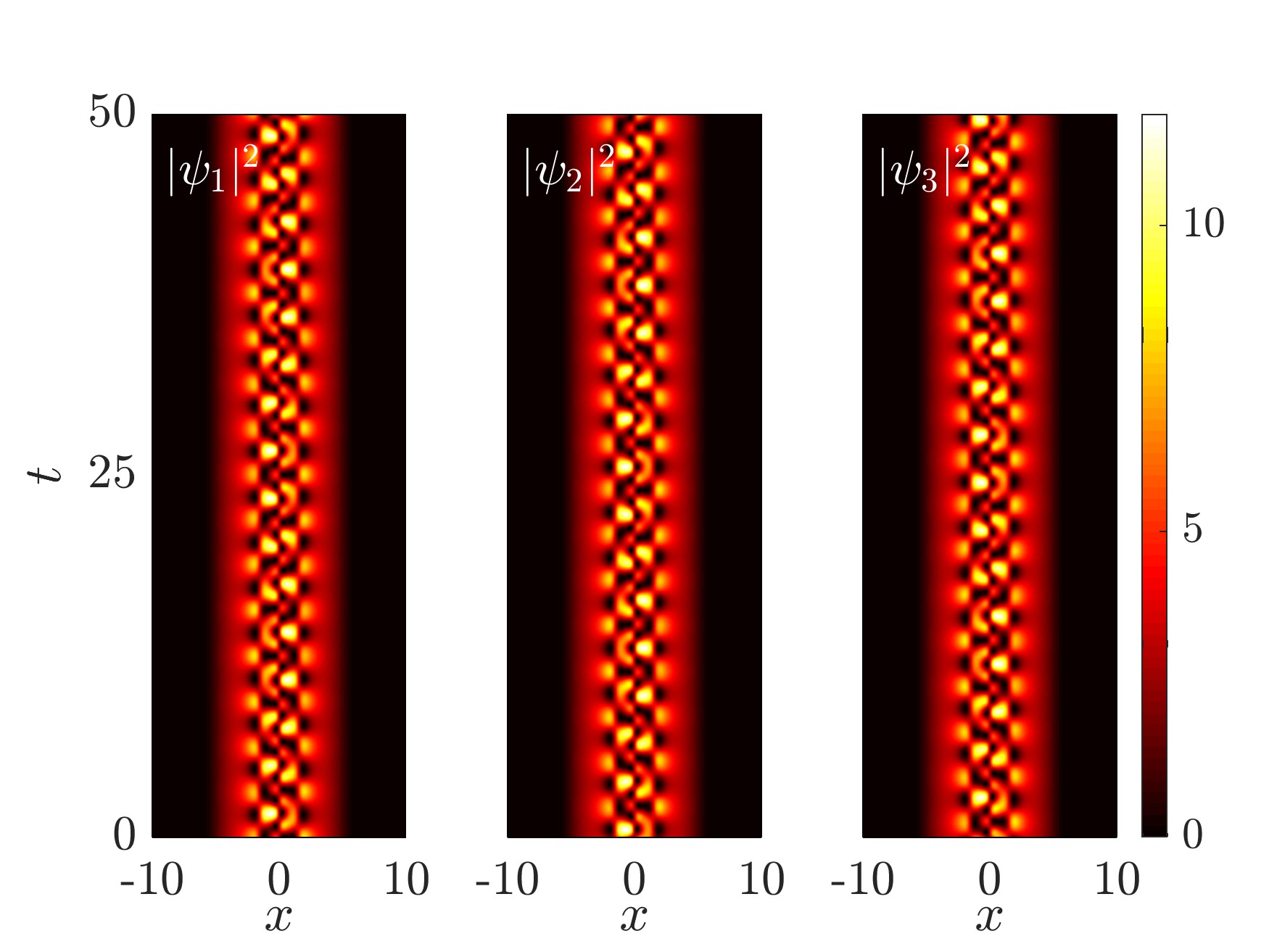}}
\subfigure[]{\includegraphics[width=0.33\textwidth]{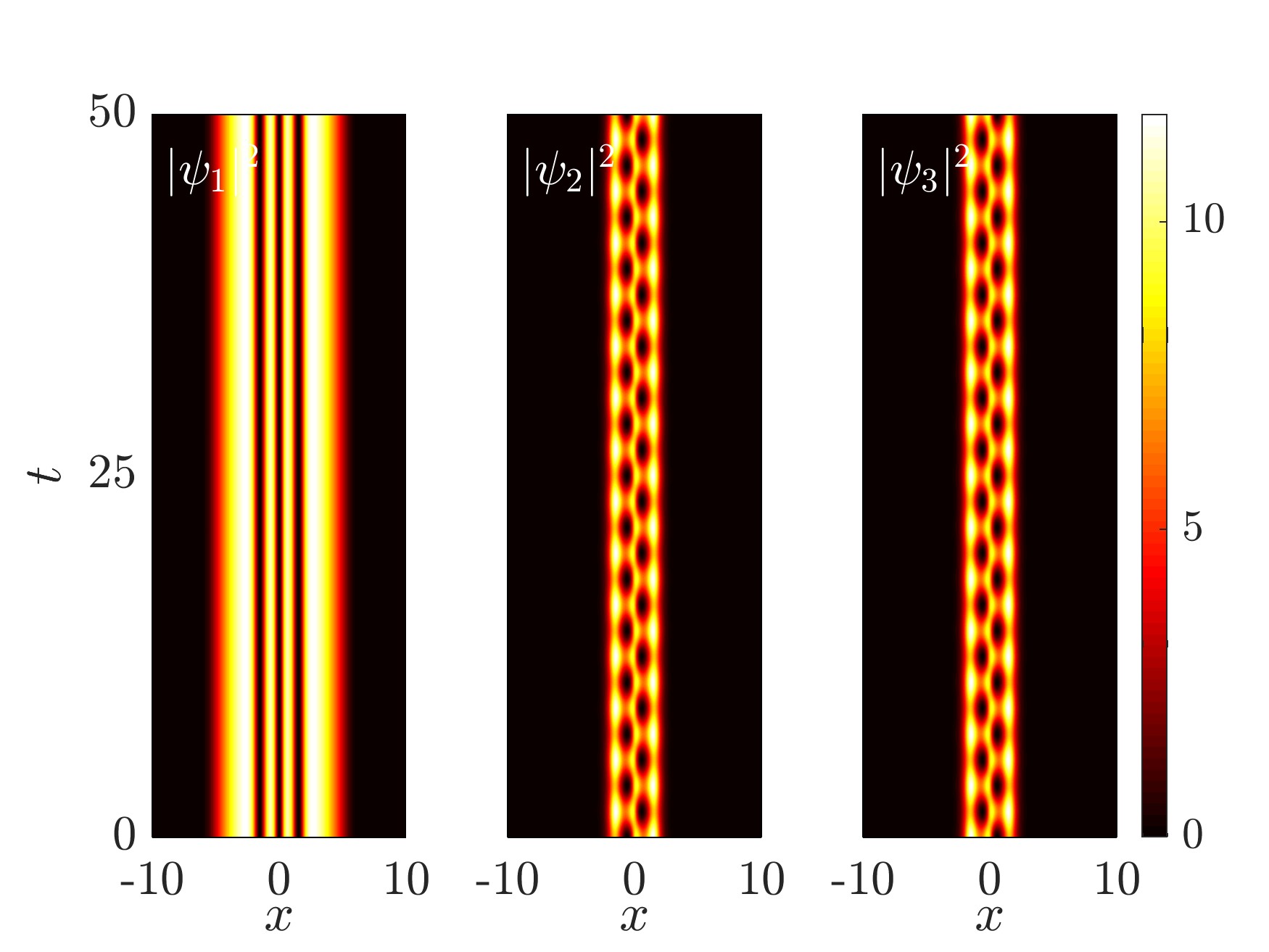}}
\subfigure[]{\includegraphics[width=0.33\textwidth]{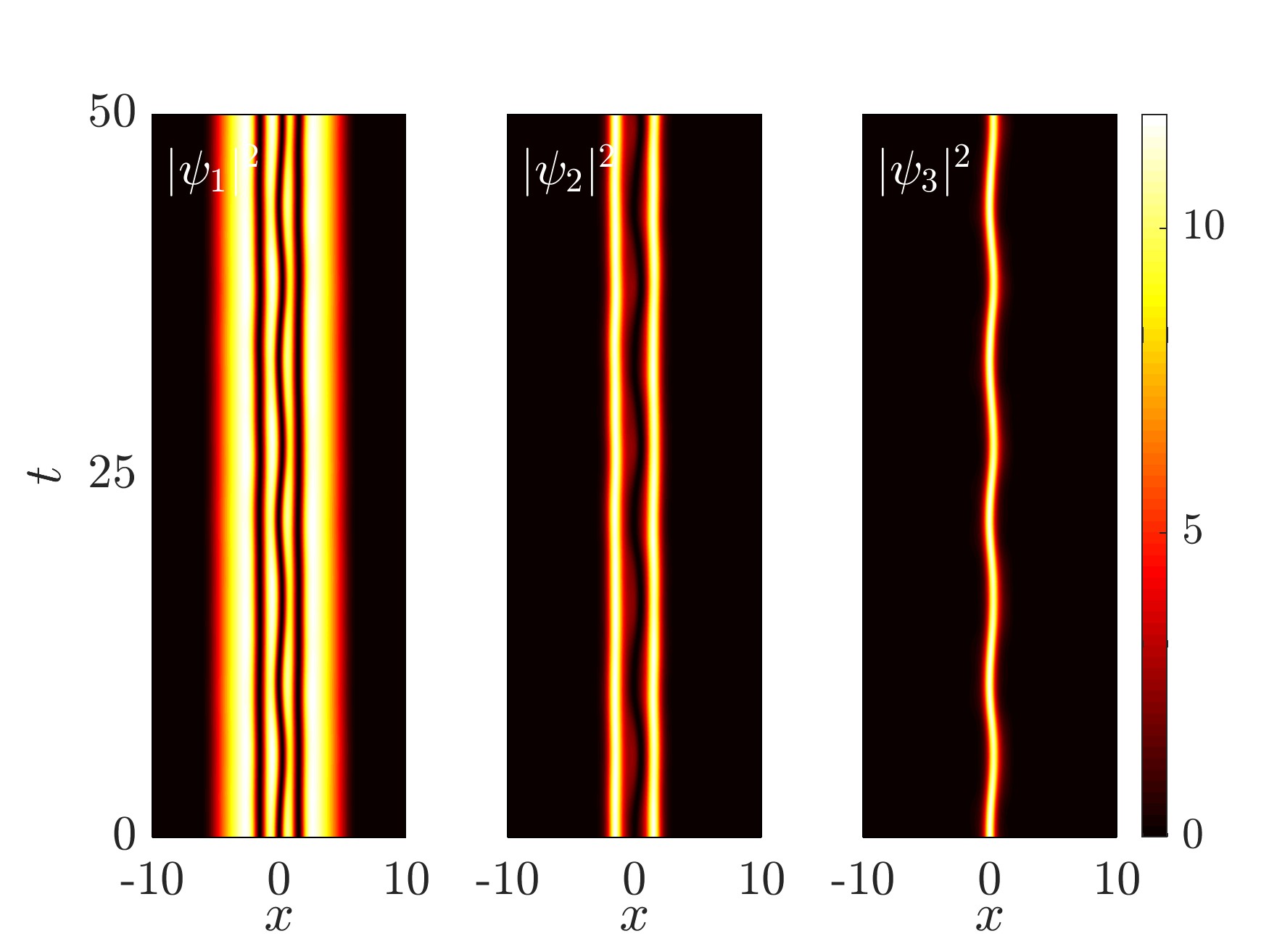}}
\subfigure[]{\includegraphics[width=0.33\textwidth]{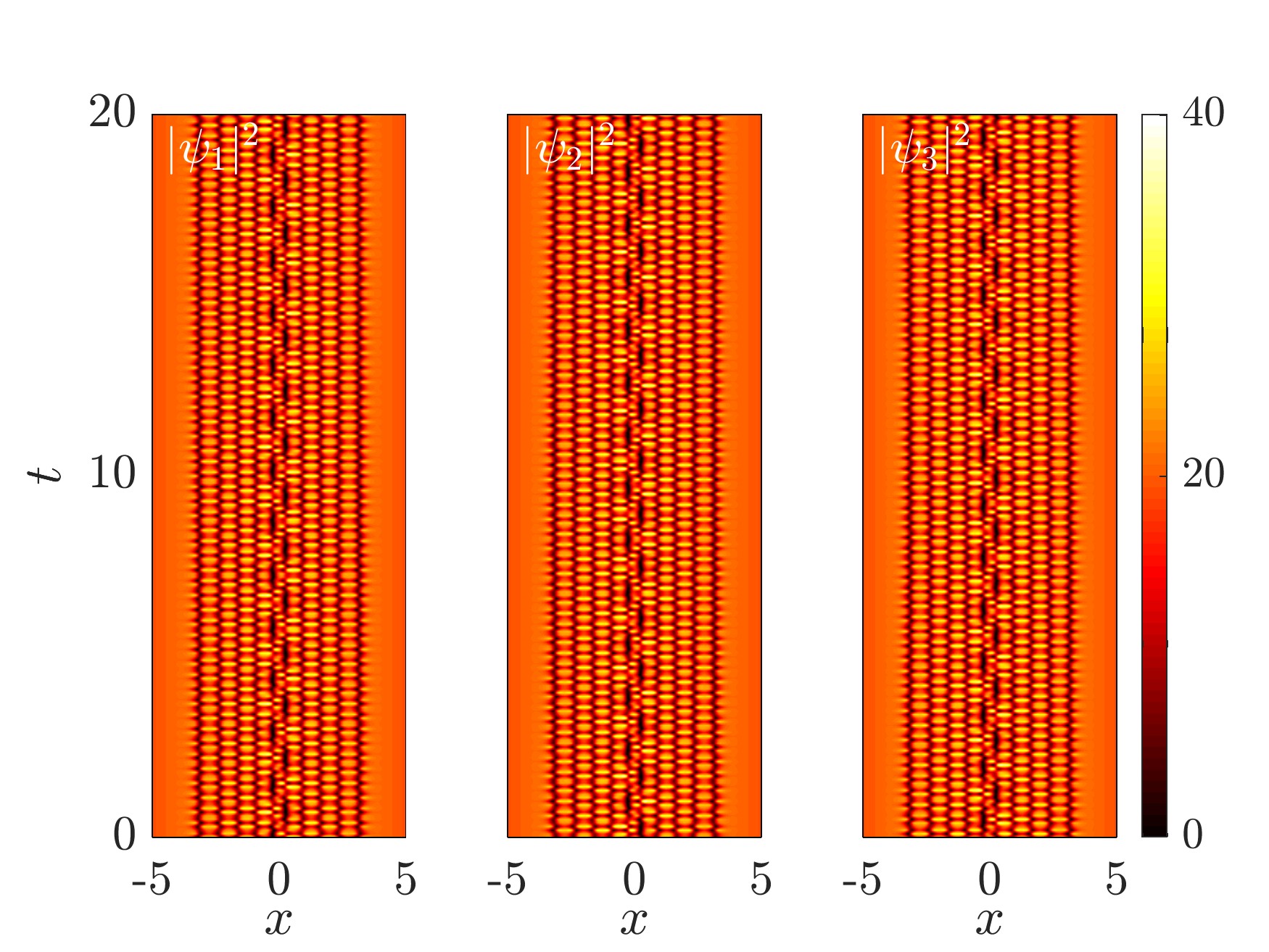}}
\subfigure[]{\includegraphics[width=0.33\textwidth]{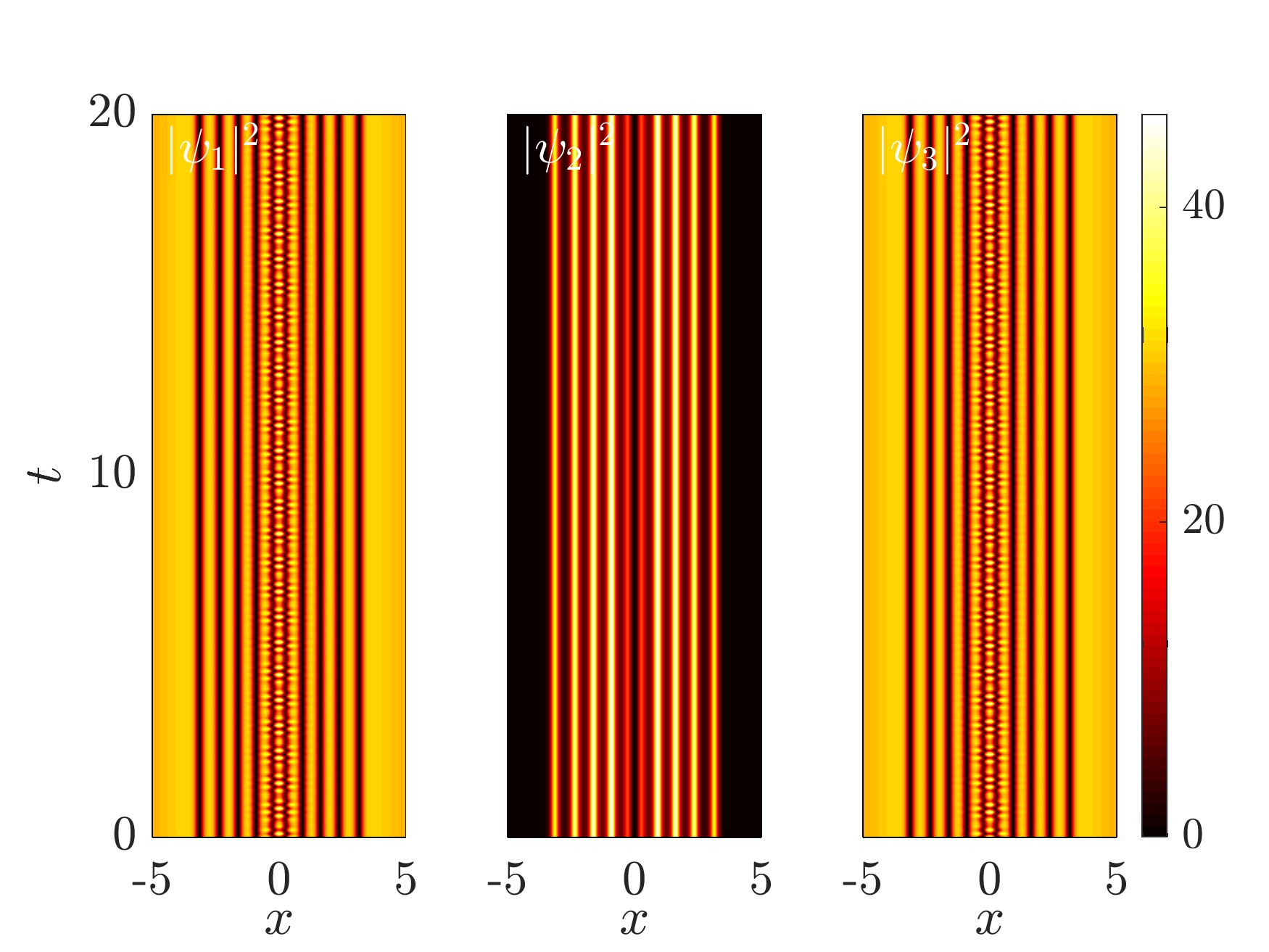}}
\subfigure[]{\includegraphics[width=0.33\textwidth]{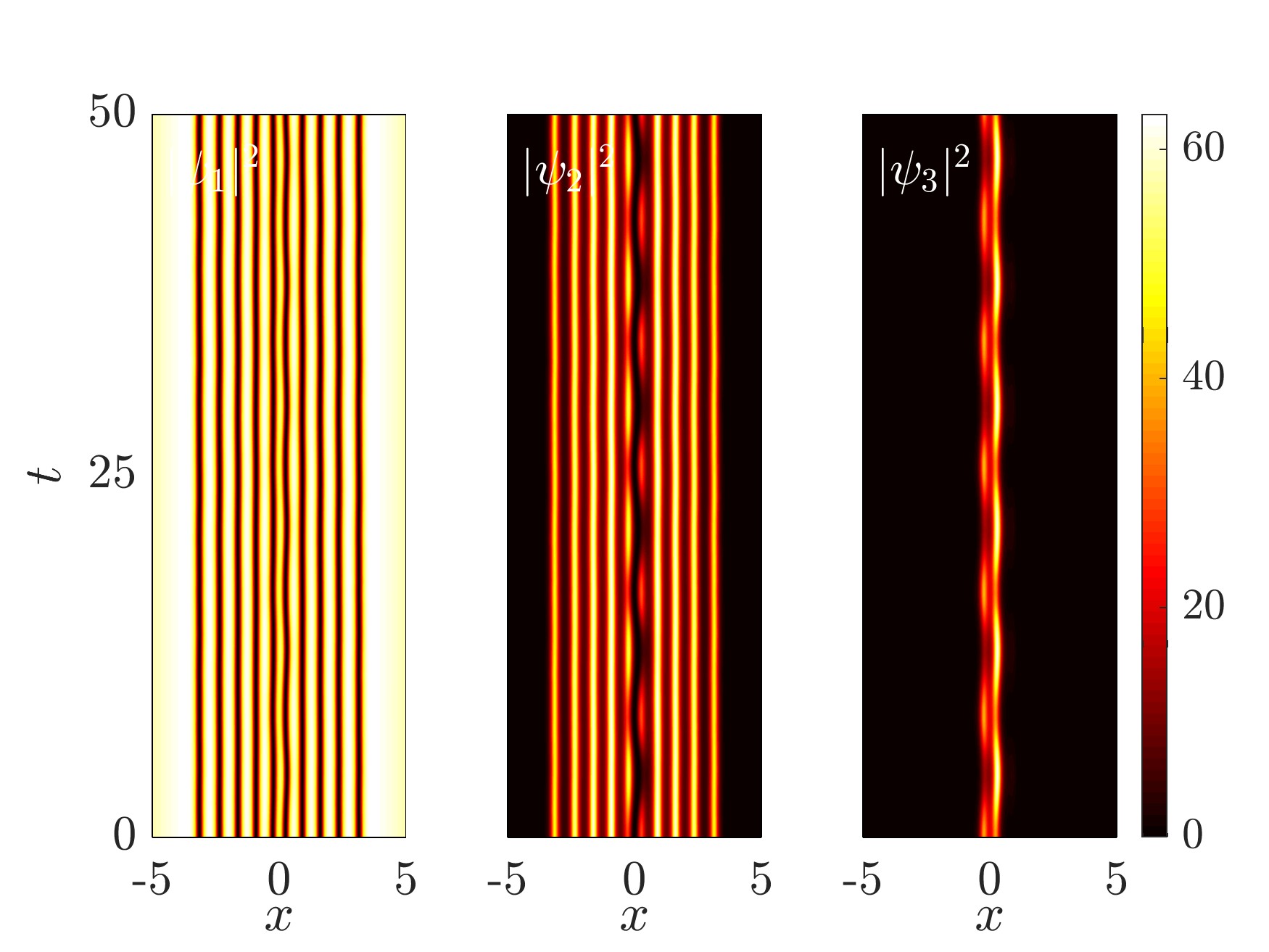}}
\caption{
\textit{Top panels: }$SU(3)$-induced dark soliton beating patterns (a) for the state \state{210} at $\mu_1=20$, each component contains two dark solitons due to the inter-component mixing. Panel (b) shows an $SU(2)$-induced out-of-phase dark-dark beating pattern, here, the first and the third components are mixed. In panel (c), we apply a driving force along the negative $x$-axis to the third component, producing the dark-bright AC oscillation in the second and third components; note that the first component is also excited. \textit{Middle panels: }The same but for the \state{310} state at $\mu_1=16$. Here, there are three solitons in each component in panel (d), and the $SU(2)$-rotation mixes the second and the third components. In panel (f), all components are excited in the AC oscillation, as the central structure is a dark-dark-bright soliton. \textit{Bottom panels: }The same but for the \state{1010} state at $\mu_1=70$. Interestingly, the rotated dynamics are also quite heterogeneous. For example in panel (h) the first and third components are mixed, the central dark solitons undergo beating dynamics while the side ones are essentially stationary.
}
\label{DynamicsS210}
\end{figure*}

We first illustrate both types of dynamics for the \state{210} state, the results are summarized in Fig.~\ref{DynamicsS210}. The first panel illustrates an $SU(3)$-induced beating pattern, and each component contains two dark solitons. This dynamics is coincidently periodic as the chemical potentials here satisfy $\mu_1-\mu_2=\mu_2-\mu_3$. In more general settings the dynamics would not be periodic, such as the $SU(3)$ beating pattern of the \state{310} state shown below, which at least has a much longer period. The second panel shows a subspace $SU(2)$-induced beating pattern mixing the first and the third components, producing the out-of-phase two dark-dark beating pattern. It is interesting that the second component sits exactly still while the other two components are very dynamical. The third panel shows a driving-induced oscillation. Here, we apply a force $F=0.1$ to the third component along the negative $x$-axis. Note that this component is somewhat locally trapped by the dark soliton of the second component. This is like a dark-bright structure (focusing on the latter two components), upon driving, due to the negative mass of the dark soliton, the central soliton propagates against the driving potential. It then deaccelerates, stopped by the driving potential, and then reverses its motion and finally closes a cycle. This is similar to the two-component dark-bright AC oscillation in \cite{Lichen:DB} but with simple Manakov interactions in our setting. Here, the dynamics is more complicated due to the presence of the first component. The two dark solitons therein are also slightly excited, e.g., its out-of-phase oscillation mode. The bright component dynamics is slightly ``rugged'' as it is influenced by the central mass of the first component, note that this is particularly the case when the bright soliton oscillates back to the trap center. Nevertheless, the prominent oscillation in the latter two components is pretty robust, and the solitary wave structures are well preserved.

\begin{figure*}
\subfigure[]{\includegraphics[width=0.24\textwidth]{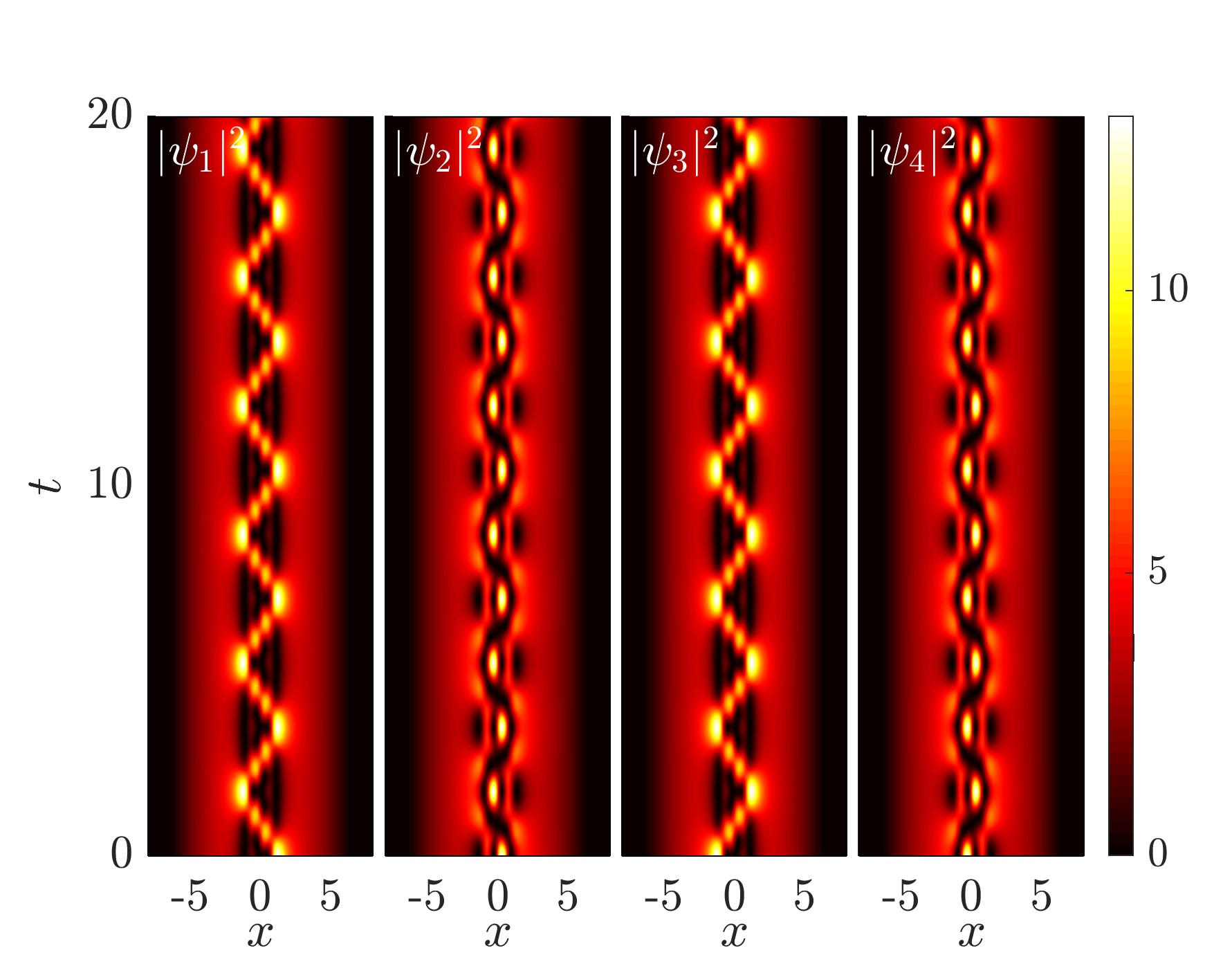}}
\subfigure[]{\includegraphics[width=0.24\textwidth]{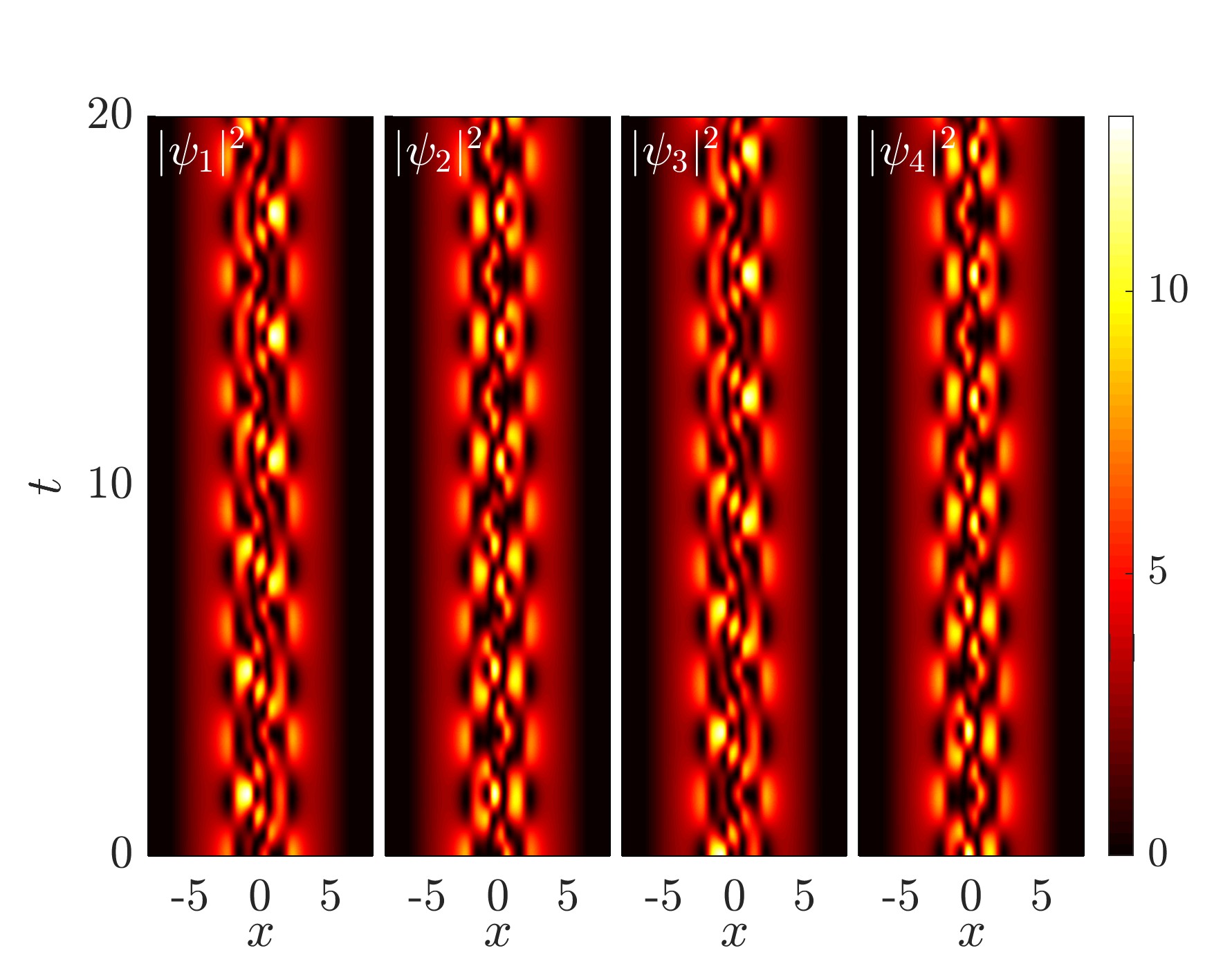}}
\subfigure[]{\includegraphics[width=0.24\textwidth]{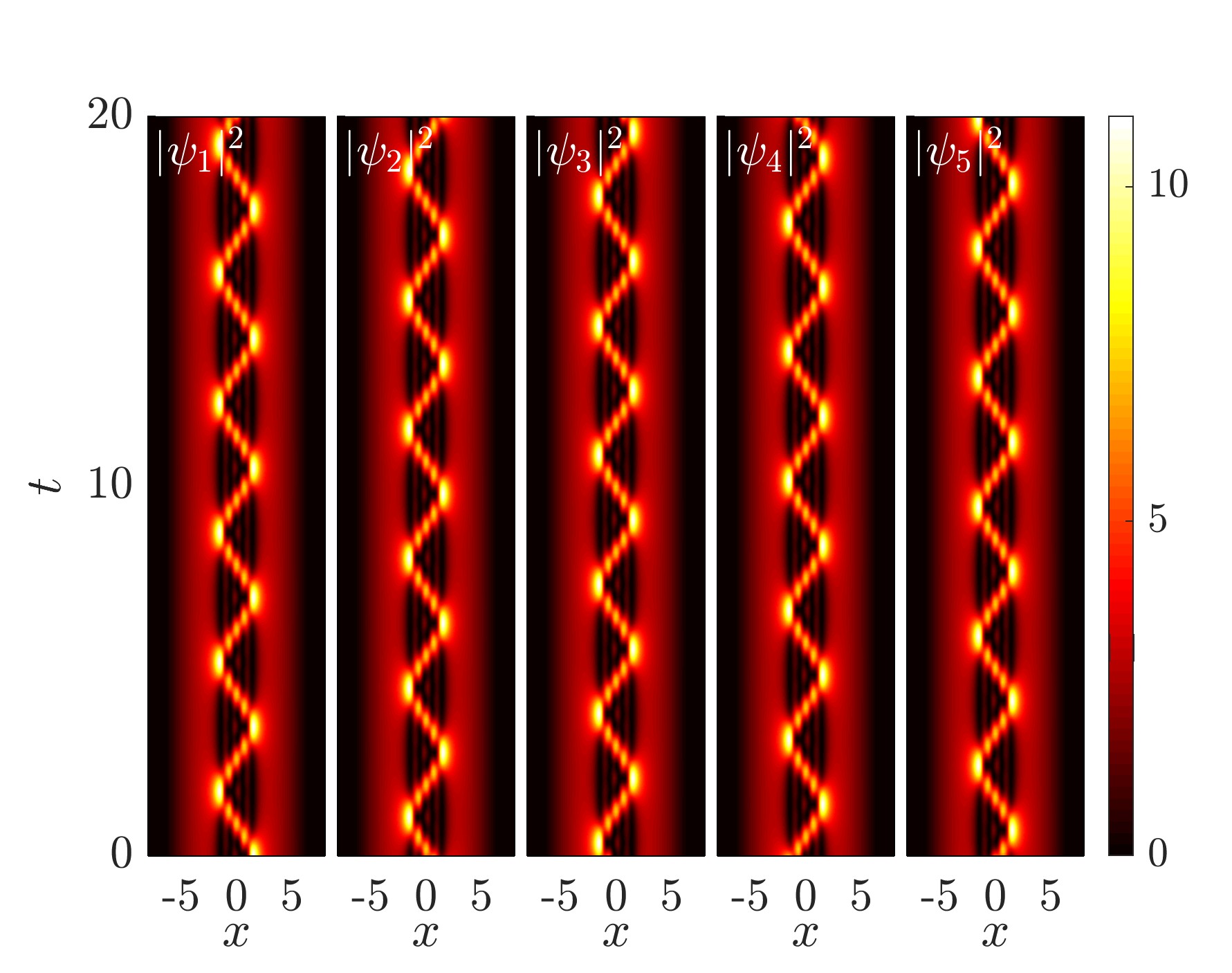}}
\subfigure[]{\includegraphics[width=0.24\textwidth]{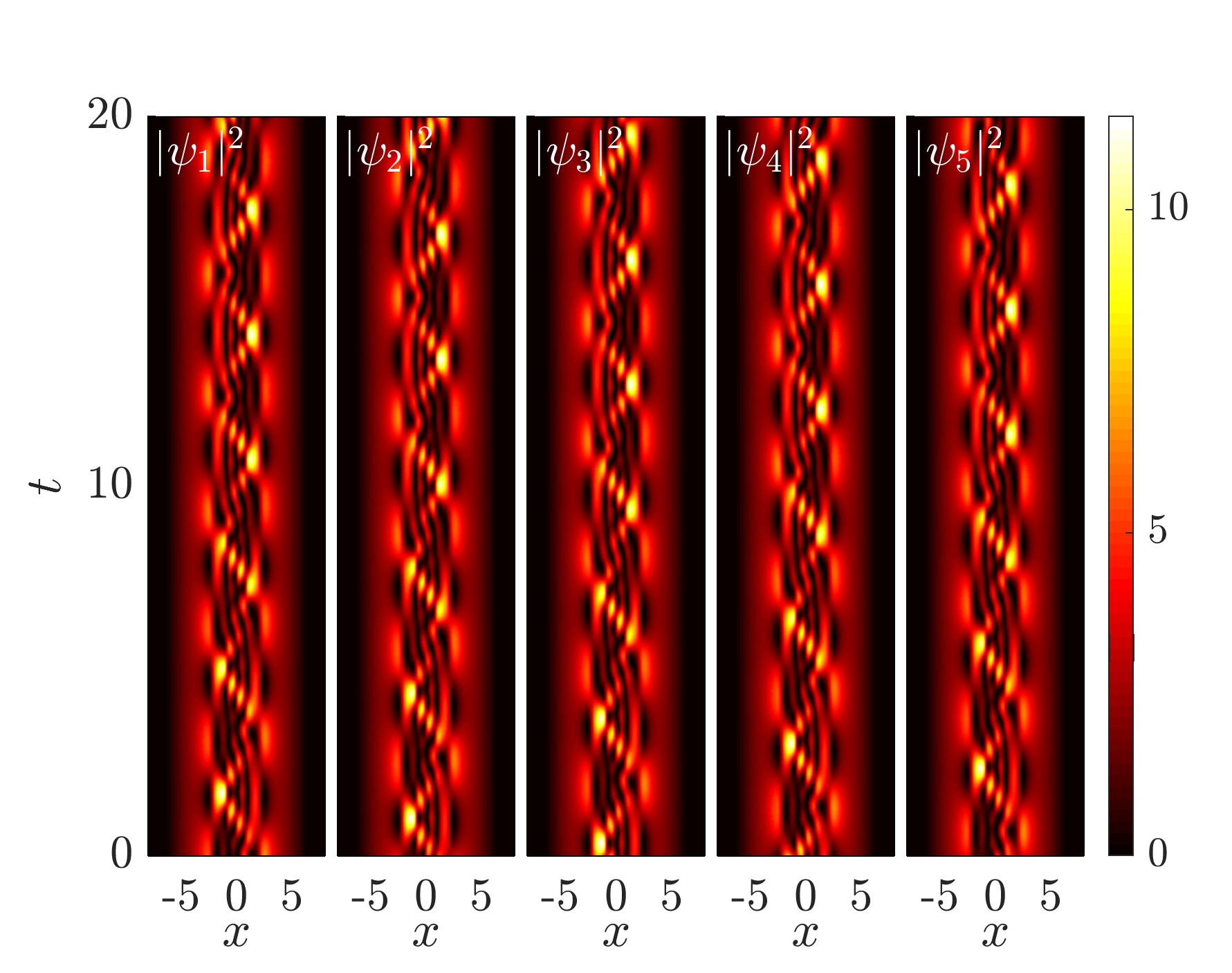}}
\caption{
\textit{Left panels:} $SU(4)$-induced beating patterns for the states \state{3210} $(a)$ and \state{4210} $(b)$ at $\mu_1=20$, each component contains three and four dark solitons, respectively, due to the inter-component mixing. 
\textit{Right panels:} $SU(5)$-induced beating patterns for the states \state{43210} $(c)$ and \state{53210} $(d)$ at $\mu_1=20$, each component contains four and five dark solitons, respectively. The $(a)$ and $(c)$ dynamics are periodic due to the special chemical potentials, the $(b)$ and $(d)$ dynamics at least have much longer periods.
}
\label{SU4}
\end{figure*}

The second row shows the same dynamics but for the \state{310} state. The $SU(3)$-induced beating pattern has three dark solitons in each component, and as mentioned above, the dynamics is no longer periodic (at least has a much longer period) upon a close inspection. The subspace $SU(2)$ rotation is applied to the second and third components, producing a simple dark-dark beating pattern. Similarly, the first component stands completely still while the other two components undergo dynamics. The driving-induced AC oscillation is much cleaner, and all the components are genuinely and coherently excited compared with that of \state{210}. This is because the central structure is a dark-dark-bright structure rather than a dark-bright-bright structure as in the \state{210} state. It is easier to balance when two dark solitons are trapping a bright soliton than when one dark soliton traps two bright solitons when one bright soliton is driven. 

The third row is for the \state{1010} state. In the $SU(3)$-induced beating pattern, there are $10$ dark solitons in each component. Interestingly, the central breathing dynamics differs from that of the sides. This is also reflected in the subspace $SU(2)$ rotation-induced dynamics in panel (h) where the central dark solitons breath while the side ones are essentially stationary in the first and third components. These are clearly consequences of the heterogeneous structure of the \state{1010} state itself. The final panel (i) shows the driving-induced AC oscillation, the dynamics is more complex as there are multiple dark and bright solitons involved, e.g., the two bright peaks are clearly not symmetric, note that the peak density on the right is much higher. Nevertheless, the oscillation remains robust despite that the structure is much more complex.

Finally, two examples are illustrated for each $SU(4)$- and $SU(5)$-induced beating dynamics in Fig.~\ref{SU4}. These patterns are pretty complicated, and the dark soliton velocities are highly asynchronized, i.e., they do not reach their minimum or maximum speeds simultaneously. Note that these patterns are already obtained from very symmetric rotations, if the rotation was less symmetric, the patterns would become even more complicated. Here, the beating patterns again can be either periodic or aperiodic depending on the specific chemical potentials. It is also possible to study subspace rotations and driving-induced dynamics and so on, and we shall not investigate these further here.

\section{Conclusions and Future Challenges}
\label{conclusion}

In this work, we presented a systematic construction of stationary vector solitary waves from their linear limits in three-, four-, and five-component Bose-Einstein condensates with repulsive Manakov interactions. We reveal their waveforms and also find suitable chemical potential intervals where they can be fully stabilized. Their waveforms are much more complex than the one- and two-component counterparts, e.g., heterogeneous lattices, and the number of them also grows much faster with respect to the principle quantum number. Some $SU(n)$-induced and driving-induced dynamics producing dark soliton oscillation patterns are also illustrated. These robust and rich structures and their versatile dynamics are ideal for future theoretical investigation and experimental implementation.

Our work demonstrates the effectiveness of the method of constructing solitary waves from the linear limit. The work can be extended in various directions, even in the present one-dimensional setting. First, studying cases away from the Manakov limit is interesting, one can look for, e.g., dark-anti-dark states \cite{engels20}. Second, it is also interesting to study the cases where the masses of the two components are different. Here, a state with a smaller quantum number may trap a state with a larger quantum number depending on the mass ratio \cite{EGC:dispersion}. Third, it is highly interesting to include also the effect of the spin-dependent interactions.

Another interesting research track is to continue from the integrable analytical limit \cite{DT,Dressingmethod,Hirota,Lakshman}. Here, asymmetric stationary states are available. If a state settles to constants in the $\pm \infty$ limits, we can gradually turn on a strong but finite box potential to render the condensates finite, and then continue the state further to the harmonic trap using, e.g., an interpolation in the potentials. It should be noted that parametric continuation can start from any analytically tractable limit, there is no reason the linear limit should be the starting point. This may generate asymmetric solitary waves in the harmonic potential.

The work should be naturally extended to higher dimensions, where the method is much more versatile because of the emergence of degenerate states at the linear limit and the additional freedom of asymmetric traps. It should be noted that there has already been a number of such studies on particular states, including quite complicated multiple vortex ring structures in three dimensions \cite{Wang:VR}. Here, our focus is different. The question is not how to construct a particular state based on physical insight, but rather the very starting point is the linear states themselves and we aim to look for well-defined rules to construct solitary waves systematically, regardless what comes out, like what we are doing here. Obviously, each linear state is a good candidate for continuation. Next, mixing degenerate states may produce novel solitonic structures, this is very different from the one-dimension setting as bound states therein cannot degenerate. For example in a two-dimensional symmetric trap, states $|n_xn_y\rangle=|10\rangle$ and $|01\rangle$ are degenerate, each produces a dark soliton stripe state. By contrast, the $(|10\rangle \pm i|01\rangle)/\sqrt{2}$ linear states produce a single vortex \cite{Alexander2001} and anti-vortex state of unit charge, respectively. Similarly, the linear state 
$(|20\rangle+|02\rangle)/\sqrt{2}$ yields a dark soliton ring.

The first step is to find all possible sets of degenerate states, this can be readily solved and can even be visualized using the lattice planes of the quantum number ``lattice''. The nontrivial part is how to mix a set of degenerate states. Proper rules should be articulated such that one can efficiently generate as many topologically distinct solitary waves as possible yet following a simple procedure. One possibility is that for any set of degenerate states of size $k$, we can choose any $m \leq k$ states out of the set and mix them using different coefficients, e.g., $\pm 1, \pm i$. If a topologically distinct state is found, it should be continued and also added to the degenerate set. More sophisticated rules should be considered, but it is immediately clear that the method should be capable of producing a (very) large array of organized solitary waves. Indeed, even the rather simple rule above should produce a diverse array of solitary waves, despite they may not be complete. Research work along these lines are currently in progress, and will be reported in future publications.

\section*{Appendix: Linear stability analysis of the $n$-component GPE}
In this Appendix, we discuss the BdG stability analysis, and calculate the BdG matrix. First, we introduce the following perturbation Ans\"atze around the stationary states of Eq.~(\ref{ss}):
\begin{eqnarray}
\psi_{j}(x,t)=e^{-i\mu_{j}t}\Big\lbrace \psi_{j}^{0}(x) + \varepsilon
\left(a_{j}(x)e^{\lambda t}+b_{j}^{\ast}(x)e^{\lambda^{\ast}t}\right)\Big\rbrace,
\label{perturb_ansatz}
\end{eqnarray}
where $\varepsilon\ll 1$. Upon substituting Eq.~\eqref{perturb_ansatz} into the GPE of Eq.~(\ref{GPE}), we obtain 
at order $O(\varepsilon)$ an eigenvalue problem of the form:
\begin{equation}
M v = \lambda v,
\end{equation}
where $v=(a_1,b_1,...,a_n,b_n)^T$ and the matrix $M$ is given by the following compact form:
\begin{eqnarray}
M=\Lambda[D+G2 \cdot (\psi2 \psi2^{\dagger})],
\end{eqnarray}
where 
\begin{eqnarray}
\Lambda &=& \mathrm{diag}[-i,i,...,-i,i], \\
D &=& \mathrm{diag}[\mathcal{L}_1+U_1,\mathcal{L}_1+U_1,...,\mathcal{L}_j+U_j,\mathcal{L}_j+U_j,...], \nonumber \\
 \\
\mathcal{L}_j &=& - \frac{1}{2} \frac{\partial^2}{\partial x^2} +V -\mu_j, \\
U_j &=& \sum_k g_{jk}|\psi_k^0|^2, \\
G2 &=& \mathrm{kron}(G, \mathrm{ones}(2,2)), \quad G_{ij}=g_{ij}, \\
\psi2 &=& (\psi_1^0, \psi_1^{0*}, ..., \psi_n^0, \psi_n^{0*})^T.
\end{eqnarray}
Here, $C=A\cdot B$ denotes the element by element multiplication, i.e., $C_{ij}=A_{ij}B_{ij}$, $\mathrm{ones}(2,2)$ is a $2\times2$ matrix with all elements equal to $1$, the kron(A,B) operator expands the $A$ matrix, where each element $A_{ij}$ is replaced by the block matrix $A_{ij}B$. One can readily check that when $n=3$, the matrix correctly restores the BdG matrix of the three-component GPE \cite{Wang:DD}.

Finally, we compute the first $100$ lowest-lying eigenvalues in magnitude for each stationary state, which correspond to the low-lying excitation modes. When the eigenvalues have positive real parts, i.e., Re$(\lambda)>0$, the stationary state is dynamically unstable with respect to perturbations. On the other hand, if the BdG spectrum is entirely imaginary, the state is robust and is dynamically stable.

\begin{acknowledgments} 
We thank P. G. Kevrekidis and Lichen Zhao for helpful discussions. We gratefully acknowledge supports from the National Science Foundation of China under Grant No. 12004268, and the Fundamental Research Funds for the Central Universities, China. We thank the Emei cluster at Sichuan university for providing HPC resources.
\end{acknowledgments}

\bibliography{Refs}

\end{document}